\pdfoutput=1
\documentclass[sigconf]{acmart}
\makeatletter
\gdef\@copyrightpermission{
 \begin{minipage}{0.3\columnwidth}
  \href{https://creativecommons.org/licenses/by/4.0/}{\includegraphics[width=0.90\textwidth]{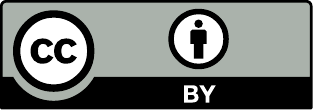}}
 \end{minipage}\hfill
 \begin{minipage}{0.7\columnwidth}
  \href{https://creativecommons.org/licenses/by/4.0/}{This work is licensed under a Creative Commons Attribution International 4.0 License.}
 \end{minipage}
 \vspace{5pt}
}
\def\@ACM@checkaffil{
    \if@ACM@instpresent\else
    \ClassWarningNoLine{\@classname}{No institution present for an affiliation}%
    \fi
    \if@ACM@citypresent\else
    \ClassWarningNoLine{\@classname}{No city present for an affiliation}%
    \fi
    \if@ACM@countrypresent\else
        \ClassWarningNoLine{\@classname}{No country present for an affiliation}%
    \fi
}
\makeatother
\AtBeginDocument{%
  }

\usepackage{multirow}
\usepackage{bm}
\usepackage{threeparttable}
\usepackage{enumitem}
\usepackage{algpseudocode}
\usepackage{algorithmicx,algorithm}

\usepackage{graphicx}

\usepackage{balance}

\usepackage{etoolbox}

\copyrightyear{2023}
\acmYear{2023}
\setcopyright{rightsretained}
\acmConference[KDD '23]{Proceedings of the 29th ACM SIGKDD Conference on Knowledge Discovery and Data Mining}{August 6--10, 2023}{Long Beach, CA, USA}
\acmBooktitle{Proceedings of the 29th ACM SIGKDD Conference on Knowledge Discovery and Data Mining (KDD '23), August 6--10, 2023, Long Beach, CA, USA}
\acmDOI{10.1145/3580305.3599319}
\acmISBN{979-8-4007-0103-0/23/08}

\begin{document}

\title{DyTed: Disentangled Representation Learning for Discrete-time Dynamic Graph}

\author{Kaike Zhang}
\affiliation{%
  \institution{Institute of Computing Technology, Chinese Academy of Sciences}
}
\affiliation{%
  \institution{The University of Chinese Academy}
  \country{of Sciences, Beijing, China}
}
\email{kaikezhang21s@ict.ac.cn}

\author{Qi Cao}
\affiliation{%
  \institution{Institute of Computing Technology, Chinese Academy of Sciences}
  \country{Beijing, China}
}
\email{caoqi@ict.ac.cn}
\authornote{Corresponding authors}

\author{Gaolin Fang}
\affiliation{%
  \institution{Tencent Inc.}
  \country{Beijing, China}
}
\email{glfang@foxmail.com}

\author{Bingbing Xu}
\affiliation{%
  \institution{Institute of Computing Technology, Chinese Academy of Sciences}
  \country{Beijing, China}
}
\email{xubingbing@ict.ac.cn}

\author{Hongjian Zou}
\affiliation{%
  \institution{Tencent Inc.}
  \country{Beijing, China}
}
\email{hongjianzou@tencent.com}

\author{Huawei Shen}
\authornotemark[1]
\affiliation{%
  \institution{Institute of Computing Technology, Chinese Academy of Sciences}
  \country{Beijing, China}
}
\email{shenhuawei@ict.ac.cn}

\author{Xueqi Cheng}
\affiliation{%
  \institution{Institute of Computing Technology, Chinese Academy of Sciences}
  \country{Beijing, China}
}
\email{cxq@ict.ac.cn}

\renewcommand{\shortauthors}{Kaike Zhang et al.}

\begin{abstract}
  Unsupervised representation learning for dynamic graphs has attracted a lot of research attention in recent years. Compared with static graph, the dynamic graph is a comprehensive embodiment of both the intrinsic stable characteristics of nodes and the time-related dynamic preference. However, existing methods generally mix these two types of information into a single representation space, which may lead to poor explanation, less robustness, and a limited ability when applied to different downstream tasks. To solve the above problems, in this paper, we propose a novel \emph{disen\textbf{T}angl\textbf{ed} representation learning framework for discrete-time \textbf{Dy}namic graphs}, namely \emph{\textbf{DyTed}}. We specially design a temporal-clips contrastive learning task together with a structure contrastive learning to effectively identify the time-invariant and time-varying representations respectively. To further enhance the disentanglement of these two types of representation, we propose a disentanglement-aware discriminator under an adversarial learning framework from the perspective of information theory. Extensive experiments on Tencent and five commonly used public datasets demonstrate that DyTed, as a general framework that can be applied to existing methods, achieves state-of-the-art performance on various downstream tasks, as well as be more robust against noise.

\end{abstract}

\ccsdesc[500]{Information systems~Social networks}
\ccsdesc[500]{Computing methodologies~Learning latent representations}

\keywords{Dynamic Graphs, Disentangled Representation Learning}

\maketitle

\section{INTRODUCTION}
Graph data, which captures the relationships or interactions between entities, is ubiquitous in the real world, e.g., social networks~\cite{liu2019characterizing}, citation graphs~\cite{10.5555/3367471.3367648}, traffic networks~\cite{Li2022TKDD}, etc. With the abundance of graph data but the expensiveness of training labels, unsupervised graph representation learning has attracted much research attention ~\cite{10.1145/2806416.2806512, zhu2020deep, cen2019anae}. It aims to learn a low-dimensional representation of each node in graphs~\cite{Xue2022DynamicsSurvey}, which can be used for various downstream tasks, including node classification and link prediction. Traditional graph representation learning mainly focuses on static graphs with a fixed set of nodes and edges~\cite{Cui2019StaticSurvey}. However, real-world graphs generally evolve, where graph structures are dynamically changing with time~\cite{sankar2020dysat}. How to learn dynamic graph representation becomes a significant research problem.

\begin{figure}
    \centering
    \includegraphics[width=3.35in]{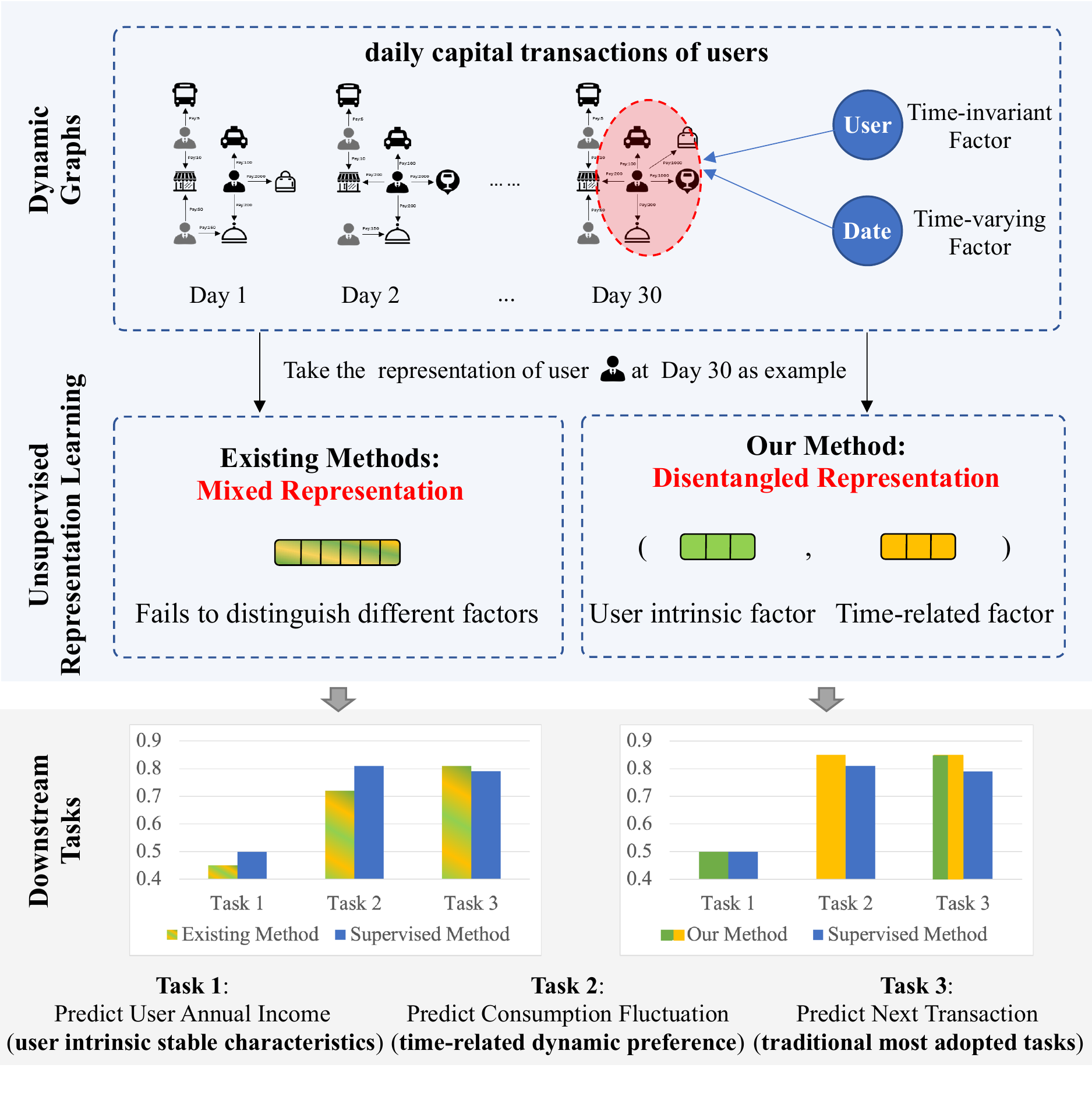}
    \caption{The effectiveness of the disentangled dynamic graphs representation in various downstream tasks. Note that, the existing method, supervised model, and our method share the same backbone model.}
    \label{fig:motivation}
\end{figure}

Existing methods for dynamic graph representation learning mainly fall into two categories~\cite{sankar2020dysat, yang2021discrete}: continuous-time approaches and discrete-time approaches. The former regards new nodes or edges of dynamic graphs in a streaming manner and models the continuous temporal information via point process~\cite{trivedi2019dyrep, zuo2018embedding} or temporal random walks~\cite{yu2018netwalk, nguyen2018dynamic}. This type of method usually requires finer-grained timestamps, which is however difficult to obtain in the real world for privacy protection.
The latter regards the dynamic graphs as a series of snapshots changing at discrete time and adopts structure models (e.g., graph neural networks~\cite{kipf2016semi, velickovic2017graph}) to capture graph characteristics and temporal models (e.g., deep sequential neural networks~\cite{bai2018empirical, vaswani2017attention}) to summarize historical information of each node~\cite{pareja2020evolvegcn, sankar2020dysat, yang2021discrete}, which is more practical.

Despite the preliminary success, existing methods typically adhere to a paradigm that generates a mixed representation for each node, neglecting to differentiate between the varying factors that determine dynamic graphs. As shown in Figure~\ref{fig:motivation}, taking the real dynamic graph of daily capital transactions on Tencent\footnote{Tencent is a Chinese internet and technology company involving social services and financial business (https://www.tencent.com/en-us/)} as an example, the daily transactions are determined by both the factor of user intrinsic characteristics and the factor of dates. Mixing these two kinds of factors into a single representation as existing methods leads to a limited ability when applied to various downstream tasks that require different types of information (bottom of Figure~\ref{fig:motivation}). In other words, there is still a lack of an effective dynamic graph representation learning method to handle various downstream tasks.

Recently, disentangled representation learning in various fields has demonstrated that separating the informative factors in representations is an essential step toward better representation learning~\cite{pmlr-v97-locatello19a,NEURIPS2021_e242660d,zhu2021and}. However, due to the non-Euclidean characteristics of graph structure and the complexity of temporal evolution, as well as the lack of guidance, how to learn disentangled dynamic graph representation still remains unexplored and challenging.

In this paper, we introduce a novel \emph{disen\textbf{T}angl\textbf{ed} representation learning framework for discrete-time \textbf{Dy}namic graphs}, namely \emph{\textbf{DyTed}}. 
Without loss of generality, we assume that the dynamic graph is a comprehensive embodiment of both the intrinsic stable characteristics of the node, referred to as the time-invariant factor, and the time-related dynamic preference, referred to as the time-varying factor. 
To effectively identify such two types of factors or information, we propose a time-invariant representation generator with the carefully designed temporal-clips contrastive learning task, together with a time-varying representation generator with structure contrastive learning. 
To further enhance the disentanglement or separation between time-invariant and time-varying representation, we propose a disentanglement-aware discriminator under an adversarial learning framework from the perspective of information theory. As shown in Figure~\ref{fig:motivation}, the different parts of disentangled representation perform well on downstream tasks, which is comparable to or even better than the supervised method. 

Extensive experiments on Tencent and five commonly used public datasets demonstrate that our framework, as a general framework that can be applied to existing methods, significantly improves the performance of state-of-the-art methods on various downstream tasks including node classifications and link predictions. Also, we offer ablation studies to evaluate the importance of each part and conduct noise experiments to demonstrate the model's robustness.

In summary, the main contributions are as follows:
\begin{itemize}[leftmargin=*]
    \item To the best of our knowledge, we are the first to study and introduce the disentangled representation learning framework for discrete-time dynamic graphs.
    \item We propose two representation generators with carefully designed pretext tasks and a disentanglement-aware discriminator under an adversarial learning framework.
    \item We conduct extensive experiments on real dynamic graphs of daily capital transactions on Tencent, achieving state-of-the-art performance on various downstream tasks.
\end{itemize}

\section{RELATED WORK}
This section briefly reviews the research on dynamic graph representation learning and disentangled representation learning.

\subsection{Dynamic Graph Representation Learning}
Representation learning for dynamic graphs aims to learn time-dependent low-dimensional representations of nodes~\cite{Xue2022DynamicsSurvey}, which can be mainly divided into continuous-time and discrete-time approaches according to the form of dynamic graphs.
Continuous-time approaches treat the dynamic graphs as a flow of nodes or edges annotated with a specific timestamp~\cite{Nguyen2018Continous, xu2020inductive, kumar2019predicting}. To incorporate the temporal information, either temporal random walks are sampled to serve as the context information of nodes~\cite{nguyen2018dynamic, yu2018netwalk} or point process are adopted regarding the arrival of nodes/edges as an event~\cite{chang2020continuous, trivedi2019dyrep, zuo2018embedding, zhou2018dynamic}. 
Although continuous-time approaches have demonstrated success, practical considerations such as privacy must be considered when collecting data in real-world scenarios, which makes it difficult to acquire fine-grained timestamps.

Another line of research regards the dynamic graphs as a series of snapshots~\cite{sankar2020dysat, goyal2018dyngem, pareja2020evolvegcn}, which generally captures the characteristics of these snapshots via the structural and temporal models. Early methods adopt the matrix decomposition to capture the graph structure in each snapshot and regularize the smoothness of the representation of adjacent snapshots~\cite{ahmed2018deepeye, li2017attributed}. Unfortunately, such matrix decomposition is usually computationally complex~\cite{Xue2022DynamicsSurvey}. With the development of deep learning, graph neural networks~\cite{kipf2016semi} are adopted to capture the structural information while recurrent neural networks or transformer~\cite{vaswani2017attention} are further utilized to summarize the historical information~\cite{wang2020generic,pareja2020evolvegcn,sankar2020dysat}. 
To effectively learn the representation, pretext tasks such as structure contrast~\cite{sankar2020dysat,10.1145/3459637.3482389}, graph reconstruction~\cite{goyal2018dyngem,cai2021structural}, and link prediction~\cite{yang2021discrete} are further adopted to guide the model learning.

However, the above existing methods for dynamic graph representation learning generally mix various factors into a single representation, which makes it difficult to generalize the learned representation to different downstream tasks.

\begin{figure}[t]
    \centering
    \includegraphics[width=3.4in]{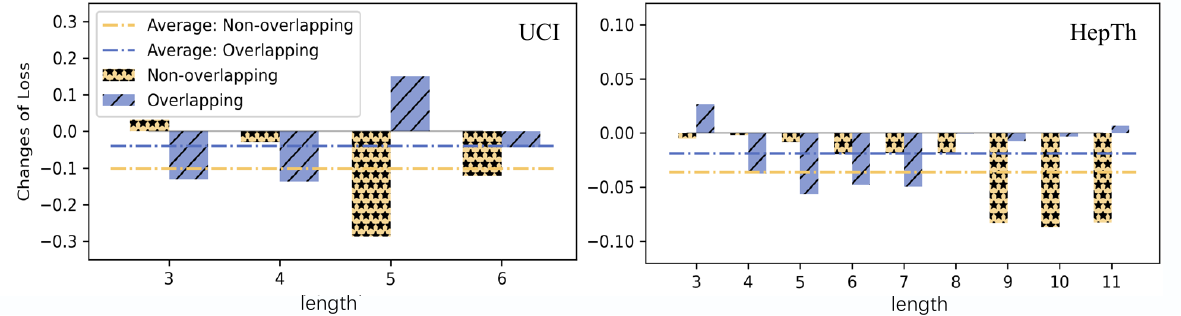}
    \caption{
        The average change proportion of overall loss after optimizing a certain type of positive pairs. 
    }
    \label{fig:smp}
\end{figure}

\begin{figure*}
    \centering
    \includegraphics[width=6.7in]{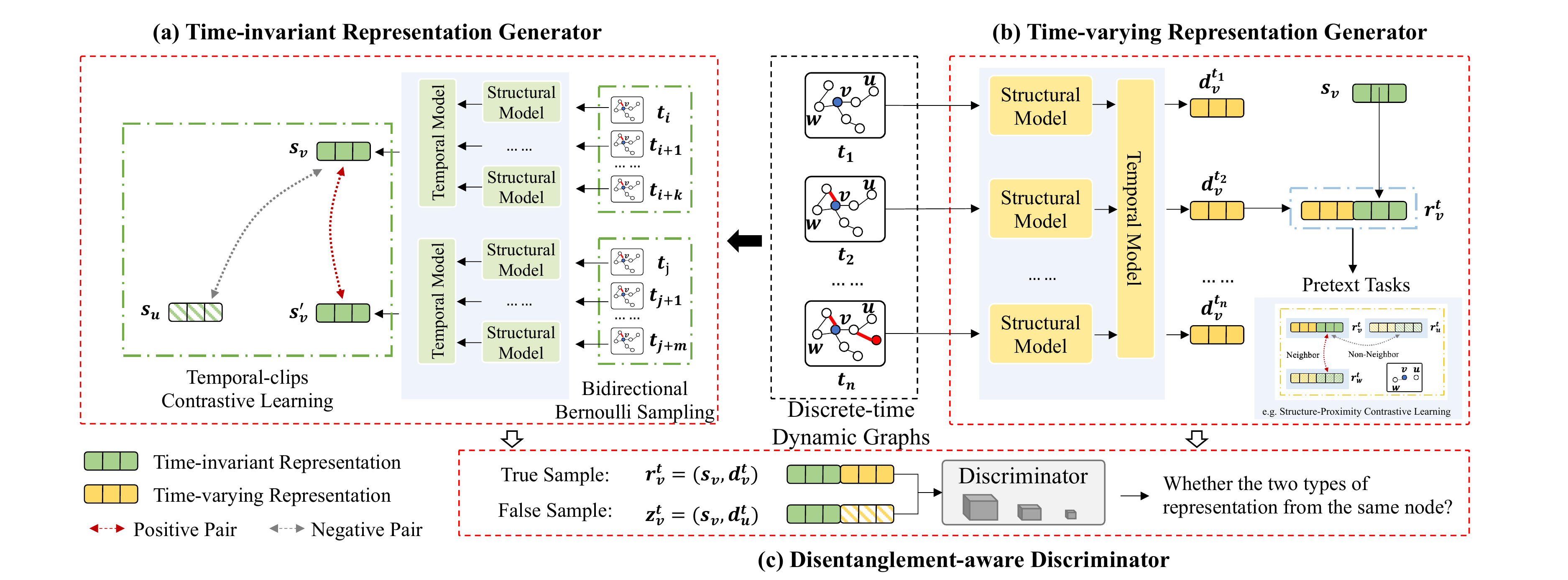}
    \caption{Overview of DyTed. (a) Time-invariant representation generator, composed of bidirectional Bernoulli sampling, structural-temporal modeling, and temporal-clips contrastive learning.
    (b) Time-varying representation generator, composed of structural-temporal modeling and pretext task in backbone models.
    (c) Disentanglement-aware discriminator, enhancing the separation between time-invariant and time-varying representation.
    }
    \label{fig:model_overview}
\end{figure*}

\subsection{Disentangled Representation Learning}
Recently, disentangled representation learning has attracted a lot of research attention and achieved great success in many fields~\cite{NEURIPS2021_e242660d,pmlr-v97-ma19a, pmlr-v97-locatello19a, Cao2022SIGIR, Zhao2022Web, wang2022disenctr}.
Specifically, in computer vision, the identity of a face is disentangled from the views or pose information to perform better on image recognition~\cite{Tran_2017_CVPR}. In natural language generation, the writing style is disentangled from the text content to serve the text-style transfer tasks~\cite{john-etal-2019-disentangled}. In graph neural networks, the factor behind the formation of each edge is disentangled for semi-supervised node classification~\cite{Zhao2022Web}.
As demonstrated in existing research, the disentangling representation is an important step toward a better representation learning~\cite{pmlr-v97-locatello19a}, which is much closer to human perception and cognition as well as can be more robust, explainable, and transferrable. 

However, due to the complexity of graph structure and temporal evolution, how to learn disentangled representation in dynamic graphs remains largely unexplored.

\section{PROBLEM DEFINITION}
In this paper, considering the availability of data, we focus on the discrete-time dynamic graph that is defined as a series of snapshots $\{\mathcal{G}^1, \mathcal{G}^2…, \mathcal{G}^T\}$, where $T$ is the total number of snapshots. 
The snapshot at time $t$, i.e., $\mathcal{G}^t = (\mathcal{V}^t, \mathcal{E}^t)$, is a  graph with a node set $\mathcal{V}^t$ and an edge set $\mathcal{E}^t \subseteq \mathcal{V}^t \times \mathcal{V}^t$. 
We use $\bm{A}^t$ to denote the adjacency matrix corresponding to the edge set $\mathcal{E}^t$. Note that, as time evolves, there may be both the appearance and disappearance of nodes or edges. 
Existing discrete-time dynamic graph representation learning aims to learn a low-dimensional representation $\bm{r}^t_v \in \mathbb{R}^d$ for each node $v \in \mathcal{V}^t$ at each timestamp $t$, which mix different types of information into a single representation space.

In this paper, we aim to disentangle the time-invariant and time-varying information in discrete-time dynamic graph representation learning, which is formally defined as follows:

\textbf{Disentangled representation learning for discrete-time dynamic graphs}  Given a dynamic graph $\{\mathcal{G}^1,\mathcal{G}^2, ..., \mathcal{G}^T \}$, for each node $v \in \mathcal{V} $, where $\mathcal{V} = \cup _{t=1}^T\mathcal{V}^t$, we aim to learn: (1) a time-invariant representation $\bm{s}_v \in \mathbb{R}^{\frac{d}{2}}$ that is independent of time and captures intrinsic stable characteristics of node $v$; (2) time-varying representations $\bm{d}_v^t \in \mathbb{R}^{\frac{d}{2}}$ at each timestamp $t$, that reflect the fluctuating preference of node $v$ related with time. The final disentangled representation $\bm{r}^t_v \in \mathbb{R}^d $ of node $v$ at timestamp $t$ is the combination of the above two types of representations:
\begin{normalsize}
\begin{math}
    \begin{aligned}
            \bm{r}_v^t = (\bm{s}_v, \bm{d}_v^t).
    \end{aligned}
\end{math}
\end{normalsize}

In comparison to traditional representation learning methods, the disentangled representation learning approach separates the time-invariant and the time-varying information to address different downstream tasks. In addition, it's worth noting that the total representation dimension of disentangled representation learning is comparable to or even less than traditional methods. Specifically, as the time-invariant representation is consistent across all snapshots, the resulting representation in disentangle representation learning is $\bm{r}_v = (\bm{s}_v, \bm{d}_v^{1,..., T}) \in \mathbb{R}^{(T+1)\times\frac{d}{2}}$, whereas the representation obtained using traditional methods is $\bm{r}_v= (\bm{r}_v^{1,..., T})\in \mathbb{R}^{T\times d}$.

\section{Method}
\label{sec:method}
In this section, we present a disentangled representation learning framework for discrete-time dynamic graphs, namely DyTed. To effectively identify the two types of information in dynamic graphs, i.e., time-invariant and time-varying factors, 
we propose two representation generators and a disentanglement-aware discriminator under an adversarial learning framework. The overview of the  framework is shown in Figure~\ref{fig:model_overview}, where the backbone model in the blue box can be replaced with any existing discrete-time dynamic graph representation learning method. Next, we introduce each part in detail.

\subsection{Time-invariant Representation Learning}
The time-invariant representation generator aims to identify the intrinsic stable characteristics of nodes, which is not easy to learn due to the lack of explicit guidance information. To address this challenge, we consider the fundamental nature behind the time-invariant representation, that is, such properties of nodes should be identified as the same in any local temporal clips. Based on this understanding, we design an effective temporal-clips contrastive learning as the pretext task, together with a bidirectional Bernoulli sampling and structural-temporal modeling module.

\subsubsection{\textbf{Observation}}
We first define the temporal clips as a part of successive snapshots. In order to achieve the above objective for time-invariant representation of node $v$, the most straightforward way is to sample all pairs of two temporal clips and optimize the representations of node $v$ in any pair to be the same. However, such a way may result in a total of $\mathcal{O}(T^4)$ pairs where $T$ is the number of snapshots, which is too large to lead to a low optimization efficiency. To solve this challenge, we aim to minimize the objective by finding some cost-effective pairs. 
Through experiments, as shown in Figure~\ref{fig:smp}, we have three interesting observations: 1). On average, optimizing non-overlapping pairs benefits more for the overall loss reduction than overlapping pairs, i.e., the yellow dotted line is lower than the blue dotted line.  2). For non-overlapping pairs, optimizing long pairs, i.e. pairs with long temporal clips, are more effective than short pairs. 3). For overlapping pairs, the result is reverse, that is, optimizing short pairs are more effective. For more details, please refer to Appendix~\ref{appendix:temporal-clips_samling}.

\subsubsection{\textbf{Bidirectional Bernoulli sampling}}
Based on the above observed three rules, we design the bidirectional Bernoulli sampling to higher the sampling probability of more cost-effective pairs of temporal clips.

\newtheorem{Definition}{Definition}[section]
\begin{Definition}
\textbf{Truncated geometric distribution}. 
Let $p \in (0, 1)$ be the probability of success on each Bernoulli trial, $m \in \mathbb{N}^+$ be the number of Bernoulli trials when we get the first success. Given $m\in [1,L]$, it follows the truncated geometric distribution as follows:

\begin{equation}
    \begin{aligned}
         Geo(p,L) \sim f(m;p;L) = p(1-p)^{(m-1)}\Phi(L)^{-1},
    \end{aligned}
\end{equation}
where $\Phi(L) = \sum^L_{m=1}p(1-p)^{(m-1)}$. 
We have $\mathbb{E}(m)=\frac{1}{p} - \frac{L(1-p)^L}{1-(1-p)^L}$.
\end{Definition}

\begin{Definition}
\textbf{Bidirectional Bernoulli sampling}. 
Let $t_i\in \mathbb{N}^+$ be the start timestamp for the first temporal clips and $t_j\in \mathbb{N}^+$ be the end timestamp for the second temporal clips. Let $L=t_j-t_i+1$ denote the clips range from $t_i$ to $t_j$. We sample $L$ from uniform distribution $U(1, T)$ and $t_i$ from $U(1, T-L)$, where $T$ is the total number of snapshots. Then the length of the two temporal clips $m_1,m_2 \in \mathbb{N}$ is drawn i.i.d. from truncated geometric distribution $Geo(p=\psi(L), L)$, where $\psi(L)$ is a decreasing function related to $L$.  
Following the above sampling process, two temporal clips are sampled as:
\begin{normalsize}
\begin{equation}
    \begin{aligned}
         \mathcal{C}_1 &= [\mathcal{G}^{t_i}, \mathcal{G}^{t_i+1},..., \mathcal{G}^{t_i+m_1-1}],\\
        \mathcal{C}_2 &= [\mathcal{G}^{t_j-m_2+1}, \mathcal{G}^{t_j-m_2+2},..., \mathcal{G}^{t_j}].
    \end{aligned}
\end{equation}
\end{normalsize}
\end{Definition}

Next, we analyze whether the pair of temporal clips $\mathcal{C}_1, \mathcal{C}_2$ sampled via the above Bidirectional Bernoulli sampling can satisfy the observed three rules.

\begin{proposition}
    Let pairs of two temporal clips $\mathcal{C}_1,\mathcal{C}_2$ be sampled from bidirectional Bernoulli sampling. Let $X=1$ denote that $\mathcal{C}_1$ and $\mathcal{C}_2$ have overlapped snapshots, otherwise $X=0$. 
    When probability $p=\psi(L)$ in truncated geometric distribution satisfies that $\frac{2}{L+2} \leq p < 1 $ and $L \geq 3$, then we have:
        \begin{equation}
    \begin{footnotesize}
            \begin{aligned}
                 \frac{Pr(X=1)}{Pr(X=0)} \leq 1, 
                 \frac{Pr(L=l+1|X=0)}{Pr(L=l|X=0)} \geq 1,
                 \frac{Pr(L=l+1|X=1)}{Pr(L=l|X=1)} \leq 1.
            \end{aligned}
    \end{footnotesize}
            \end{equation}
    \label{pro:overlap}
\end{proposition}
$Proof.$ see Appendix~\ref{proof:overlap}.
In Proposition~\ref{pro:overlap}, 
\begin{small}
    $\frac{Pr(X=1)}{Pr(X=0)} \leq 1$   
\end{small}
indicates that the sampled probability of non-overlapping pairs is higher than overlapping pairs, satisfying the \emph{observation~1)}. 
\begin{small}
    $\frac{Pr(L=l+1|X=0)}{Pr(L=l|X=0)} \geq 1$
\end{small}
indicates that for non-overlapping pairs, the sampled clips range $L$ is more like to be large. Recall that $p=\psi(L)$ is a decreasing function related to $L$, i.e., $\psi(L)' \leq 0$,  then we can easily obtain that $\nabla_L \mathbb{E}_{G(p=\psi(L), L)}(m) \geq 0$.
In other words, the larger the sampled clips gap $L$, the longer the expected length of the sampled temporal clips $\mathbb{E}(m)$ is, satisfying \emph{observation~2)}. Similarly, 
\begin{small}
  $\frac{Pr(L=l+1|X=1)}{Pr(L=l|X=1)} \leq 1$  
\end{small}
indicates the satisfaction of \emph{observation~3)}.

To sum up, as long as $\frac{2}{L+2} \leq p=\psi(L) < 1 $, $\psi(L)'\leq 0$, and $L \geq 3$, we have demonstrated that the proposed Bidirectional Bernoulli sampling has the ability to sample the cost-effective pairs of temporal clips with higher probability. In this paper, we design an exampled implementation of $\psi(L)$ to satisfy these constraints, i.e., $\psi(L) = 1 - \alpha \frac{L}{L+2}$, where the $0<\alpha \leq 1 $ is a learnable parameter.

\subsubsection{\textbf{Temporal-clips Contrastive Learning}}
With the development of deep neural networks, the state-of-the-art methods for discrete-time dynamic graph representation learning generally follow a structural-temporal paradigm~\cite{sankar2020dysat, yang2021discrete}. Here, we also take such structural-temporal models to generate the representation.
Specifically, given two sampled temporal clips $\mathcal{C}_1$ and $\mathcal{C}_2$, the representation of time-invariant generator for node $v$ is denoted as:
        \begin{equation}
        \begin{aligned}
            \bm{s}^1_v = {G_{i}} (\mathcal{C}_1)_v,      
            \bm{s}^2_v = {G_{i}} (\mathcal{C}_2)_v,
        \end{aligned}
    \end{equation}
where $G_{i}$ can be any backbone model in existing methods.

To optimize the time-invariant generator $G_{i}$, we take the representation of node $v$ for two temporal clips sampled via Bidirectional Bernoulli sampling as the positive pair in contrastive learning, i.e., $(\bm{s}_v^1,\bm{s}_v^2)$, and take the representation of node $v$ for $\mathcal{C}_1$ and the representation of node $u$ for $\mathcal{C}_1$ as the negative pair, i.e., $(\bm{s}_v^1,\bm{s}_u^1)$.
We use InfoNCE~\cite{oord2018representation} as the contrastive loss to separate the positive pair and the negative pairs. 
Such contrastive loss can maximize the mutual information between representations in positive pairs while minimizing the mutual information between representations in negative pairs, ensuring that the extracted representations of the same node in different temporal clips are similar. 
The loss function is formalized as follows:
\begin{small}
    \begin{equation}
        L_{i}(G_i) = \mathbb{E}_v \left [- \log \left (\frac{\exp \left ({\rm sim} (\bm{s}_v^1, {\bm{s}_v^2})/\tau \right )}{\sum_{u \in \mathcal{V}} \exp \left ({\rm sim} (\bm{s}_v^1, {\bm{s}_u^1} )/\tau \right)} \right ) \right ],
    \end{equation}
\end{small}
where $\rm sim(\cdot)$ is the similarity function, and $\tau$ is the temperature.

Note that, the final time-invariant representation $\bm{s}_v$ for node $v$ is obtained when the temporal clip contains all snapshots, i.e., $\bm{s}_v = {G_{i}} ([\mathcal{G}^{1}, \mathcal{G}^{2},..., \mathcal{G}^{T}])_v$.

\subsection{Time-varying Representation Learning}
For the time-varying representation generator, we use the same backbone model as the time-invariant generator with non-shared parameters. The time-varying representation for node $v$ at timestamp $t$ is denoted as $\bm{d}_v^t$:

\begin{equation}
    \begin{aligned}
        [\bm{d}_v^1,\bm{d}_v^2,...,\bm{d}_v^T] &= {G_{v}} \left ([\mathcal{G}^{1}, \mathcal{G}^{2},..., \mathcal{G}^{T}] \right )_v.
    \end{aligned}
\end{equation}

Considering that the graph structure and the evaluation are the comprehensive embodiment of both the time-invariant and time-varying representations, we combine the time-invariant representation $\bm{s}_v$ and time-varying representation $\bm{d}_v^t$ together, i.e., $\bm{r}_v^t = (\bm{s}_v,\bm{d}_v^t)$, to complete the self-supervise task related to the evolution characteristics of dynamic graphs. Worth noting, the self-supervise task can be the same as the tasks in the backbone model or other well-design self-supervising tasks, such as structure-proximity contrastive learning:

\begin{small}
\begin{equation}
    \begin{aligned}
            L_{v}(G=\{G_i, G_v\}) = \sum_{t=1}^T \mathbb{E}_{(u,v)\in \mathcal{E}_t} \left[-\log \left (\frac{\exp \left ({\rm sim}(r_v^{t}, r_u^{t})/\tau \right )}{\sum_{w \in \mathcal{U}} \exp \left ({\rm sim}(r_v^{t}, r_w^{t})/\tau \right )} \right ) \right ].
    \end{aligned}
\end{equation}
\end{small}
or link prediction:
\begin{small}
\begin{equation}
    \begin{aligned}
             L_{v}(G) = -\sum_{t=1}^{T-1} \left [\mathbb{E}_{(u,v)\in \mathcal{E}_{t+1}} \log \left (r_v^{t\top}r_u^{t} \right ) + \mathbb{E}_{(u,w)\notin \mathcal{E}_{t+1}}  \log \left (1-(r_v^{t\top}r_w^{t}) \right ) \right].
    \end{aligned}
\end{equation}
\end{small}

\subsection{Disentanglement-aware Discriminator}
\renewcommand{\algorithmicrequire}{ \textbf{Input:}}     
\renewcommand{\algorithmicensure}{ \textbf{Output:}}    
\begin{small}
\begin{algorithm}[t]
    \caption{The training procedure of DyTed} 
    \label{al:dyted}
    \begin{algorithmic}[1]
        \Require{Dynamic graph $[\mathcal{G}^1,\mathcal{G}^2, ..., \mathcal{G}^T ]$}
        \Ensure{Time-invariant representations $\bm{S} \in \mathbb{R}^{N \times \frac{d}{2}}$; Time-varying representations $\bm{D}^t \in \mathbb{R}^{N \times \frac{d}{2}}$}, $t=1,...,T$.
        \While{not Convergence}
            \State Draw $L$ from $U(1, T)$ and $t_i$ from $U(1, T-L)$, calculate $t_j = t_i + L - 1$.
            \State Calculate the sampling vectors $\bm{k}^1, \bm{k}^2$ according to Eq.~\ref{eq:sample}
            \State Calculate the snapshots index vector $\bm{k}^i=\{k^i_j=k^i_j+k^i_{j+1}\}_{j=L-1}^1$, $i=1,2$
            \State $\mathcal{C}_1 = [\mathcal{G}^{m}\cdot k^1_{m-t_i}]_{m=t_i}^{t_j}$,  $\mathcal{C}_2 = [\mathcal{G}^{m}\cdot k^2_{t_j-m}]_{m=t_i}^{t_j}$
            \State Time-invariant representations: $\bm{S}^1 = G_i(\mathcal{C}_1)$, $\bm{S}^2 = G_i(\mathcal{C}_2)$
            \State Time-varying representations: $[\bm{D}^1,\bm{D}^2,...,\bm{D}^T] = {G_{v}} ([\mathcal{G}^i]_{i=1}^T)$
            \State  Calculate $Loss(G)$ according to Eq.~\ref{eq:loss} and minimize $Loss(G)$
            \For {Epochs of Discriminator}
                \State Calculate $Loss(D)$ according to Eq.~\ref{eq:gan} and minimize $Loss(D)$
            \EndFor
        \EndWhile
        \State Final time-invariant representation: $\bm{S} = G_i([\mathcal{G}^i]_{i=1}^T$) 
        \State \Return $\bm{S}$, $[\bm{D}^1,\bm{D}^2,...,\bm{D}^T]$
    \end{algorithmic}
\end{algorithm}
\end{small}

After the time-invariant and the time-varying representation generator, we get two parts of representations. To further enhance the disentanglement between these two types of information, we propose an adversarial learning framework from the perspective of information theory.

Imagine that if the time-invariant representations $\bm{s_v} \in \mathbb{R}^{\frac{d}{2}}$ and the time-varying representations $\bm{d^t_v} \in \mathbb{R}^{\frac{d}{2}}$ are not well disentangled, i.e., there is some overlap in the information of the representations. From the perspective of information theory, this overlap can be quantified using the mutual information $I(\bm{S}, \bm{D})$, where $\bm{S}$ is the distribution of time-invariant representations for all nodes and $\bm{D}$ is the distribution of time-varying representations for all nodes. 
In other words, to enhance the disentanglement between the time-invariant and time-varying representations, we aim to minimize the mutual information between them, i.e.,
\begin{normalsize}
\begin{equation}
        \min_{G=\{G_{i},G_{v}\}}  I(\bm{S}, \bm{D}),
\label{eq:gan}
\end{equation}
\end{normalsize}

Considering that $I(\bm{S}, \bm{D}) = \sum_{s\in \bm{S}} \sum_{d\in \bm{D}}p(s,d)\log \frac{p(s,d)}{p(s)p(d)}= \mathcal{D}_{kl}(p(\bm{S}, \bm{D})||p(\bm{S})p(\bm{D}))$, minimizing the mutual information is equivalent to minimizing the KL divergence between distribution $p(\bm{S}, \bm{D})$ and $p(\bm{S})p(\bm{D})$. As a result, we construct the true data distribution as $p_{d}=p(\bm{S}, \bm{D})$ and the generated data distribution as $p_g=p(\bm{S})p(\bm{D})$, and introduce the framework of generative adversarial networks to minimize the KL divergence between these two data distributions, i.e., 
\begin{small}
\begin{equation}
        \min_{G} \max_{D} V(G,D) = \min_{G} \max_{D} \mathbb{E}_{T}  \left(\log\left(D\left(\bm{r}_i^t\right)\right) + \log\left(1-D\left(\bm{z}_i^t\right)\right)\right)
\label{eq:gan}
\end{equation}
\end{small}
where $\bm{r}_i^t=(\bm{s}_v, \bm{d}_v^t)$ is the true sample that randomly sampled from the same node, $\bm{z}_i^t=(\bm{s}_v, \bm{d}_u^t)$ is the false sample that randomly sampled from two nodes. $D$ is the disentanglement-aware discriminator implemented by multi-layer perceptron (MLP), which can distinguish whether the time-invariant and the time-varying representation come from the same node, telling generators if two parts representations have overlapped information.

Finally, the loss function of the discriminator is:
\begin{small}
  \begin{equation}
    \begin{aligned}
          Loss(D) = - V(G,D)
    \end{aligned}
  \end{equation}  
\end{small}
and the loss function of the generators is:
\begin{small}
  \begin{equation}
\label{eq:loss}
    \begin{aligned}
          Loss(G=\{G_i, G_v\}) = L_v(G) + \lambda_1 L_i(G_i) + \lambda_2 V(G,D) + \lambda_3||w||_2^2
    \end{aligned}
\end{equation}  
\end{small}
where $\lambda_1, \lambda_2$ and $\lambda_3$ are hyperparameters.

\subsection{Training Strategy}
In order to train the parameters of bidirectional Bernoulli sampling, we introduce the categorical reparameterization~\cite{jang2016categorical}.
Given the truncated geometric distribution $Geo(p,L)$ and the probabilities of sampling length $m=1,2,...,L$ for temporal clips as $p_m=f(m;p;L)$, then we can generate $L$-dimensional sampling vectors $\bm{k} \in \mathbb{R}^l$:
\begin{small}
\begin{equation}
\label{eq:sample}
    \begin{aligned}
            k_m = \frac{\exp \left ( \left ( \log(p_m) + g_m \right ) /\tau_g \right )}{\sum_{j=1}^L \exp \left ( \left (\log(p_j)+ g_j \right ) /\tau_g \right )}, ~~~~\rm{for} ~m=1,2,...,L.
    \end{aligned}
\end{equation}
\end{small}
where $g_m$ is i.i.d. drawn from Gumbel distribution $\rm{Gumbel}(0, 1)$ and $\tau_g$ is the temperature parameter.

Finally, we can propagate the gradient to parameter $\alpha$ in bidirectional Bernoulli sampling by sampling vectors $\bm{k}$.
The procedure of training DyTed is illustrated in Algorithm~\ref{al:dyted}.

\subsection{Complexity Analysis}

Without loss of generality, we assume that the time complexity of the backbone model is $\mathcal{O}(M)$. Then, when we further attach our framework to the backbone model, the time complexity is $\mathcal{O}(\frac{3}{2}M + n|\mathcal{V}|d + n'd)$, where $\frac{3}{2}$ is due to three calculations of $\mathcal{C}_1$, $\mathcal{C}_2$ in the time-invariant generator and the entire dynamic graph in the time-varying generator, as well as the half dimension of representation in DyTed. $\mathcal{O}(n|\mathcal{V}|d)$ is the time complexity of the temporal-clips contrastive learning, where $n$ is the number of negative pairs for each positive pair in InfoNCE, and $d$ is the dimension of representations.
$\mathcal{O}(n'd)$ is the time complexity of the discriminator, where $n'$ is the number of ture and false samples. In general, $\mathcal{O}(\mathcal{M}) \gg \mathcal{O}(n|\mathcal{V}|d+n'd)$. For example, when the backbone model is a simple GCN plus LSTM, $\mathcal{O}(M) = \mathcal{O} (T|\mathcal{E}||\mathcal{V}|d)\gg \mathcal{O}(n|\mathcal{V}|d+n'd)$. As a result, compared with the backbone model, the time complexity of DyTed does not increase in magnitude. See Figure~\ref{fig:running} in Appendix~\ref{appendix:runtime} for more experimental results.

\section{EXPERIMENTS}
\begin{table}[t]
  \centering
  \caption{Datasets}
    \begin{tabular*}{\hsize}{@{}@{\extracolsep{\fill}}cccc@{}}
    \toprule
    \textbf{ DATASET } & \textbf{ \#Nodes } & \textbf{ \#Edges } & \textbf{\#Snapshots}\\
    \midrule
     UCI  &           1,809  &         56,459  & 13  \\
     Bitcoin  &           3,782  &         483,700  & 20  \\
     AS733  &           4,648  &       532,230  & 30  \\
     HepTh  &           7,576  &       196,463  & 23 \\
     HepPh  &         10,404  &       339,556  & 20  \\
     Tencent-alpha  &         11,623  &       102,464  & 30   \\
     Tencent-beta  &         115,190  &       6,680,276  & 30  \\
    \bottomrule
    \end{tabular*}%
  \label{tab:datasets}%
\end{table}%

In this section, we conduct extensive experiments to answer the following research questions (\textbf{RQs}). Due to space limitations, some experimental results or details are in Appendix.
\begin{itemize}[leftmargin=*]
    \item \textbf{RQ1:} Whether the DyTed framework can improve the performance of existing methods in various downstream tasks?
    \item \textbf{RQ2.} What does each component of DyTed bring?
    \item \textbf{RQ3:} Is there any additional benefit of disentanglement?
\end{itemize}

\subsection{Experimental Setup}
\begin{table*}[t]
\centering
\caption{Node classification for Tencent-alpha}
\resizebox{\textwidth}{!}{
    \begin{threeparttable}
    \begin{tabular}{lcccccccc|cc}
    
    \toprule
    \multirow{2}[2]{*}{\textbf{Model}} & \multicolumn{2}{c}{\textbf{Annual Income}} & \multicolumn{2}{c}{\textbf{Age}}        & \multicolumn{2}{c}{\textbf{Assets}}    & \multicolumn{2}{c}{\textbf{Financing Risk}}   & \multicolumn{2}{c}{\textbf{Consumption Fluctuation}} \\
                                         \cmidrule(r){2-3}                            \cmidrule(r){4-5}                         \cmidrule(r){6-7}                        \cmidrule(r){8-9}                               \cmidrule(r){10-11}
                                       & \textbf{micro-F1} & \textbf{macro-F1}      & \textbf{micro-F1} & \textbf{macro-F1}   & \textbf{micro-F1} & \textbf{macro-F1}  & \textbf{micro-F1} & \textbf{macro-F1}         & \textbf{micro-F1} & \textbf{macro-F1}\\
    \midrule
    \textbf{LSTMGCN}                   & 50.38 $\pm$ 0.38 & 27.12 $\pm$ 0.32 & 34.86 $\pm$ 0.48 & 17.98 $\pm$ 1.08 & 24.06 $\pm$ 0.46 & 16.37 $\pm$ 0.39 & 49.13 $\pm$ 0.32 & 29.56 $\pm$ 0.64 & 93.70 $\pm$ 0.06 & 93.65 $\pm$ 0.06\\
    \textbf{LSTMGCN-DyTed}             & (+7.23\% $\uparrow$) & (+10.55\% $\uparrow$) & (+1.55\% $\uparrow$) & (+23.30\% $\uparrow$) & (+6.77\% $\uparrow$) & (+20.95\% $\uparrow$) & (+2.89\% $\uparrow$) & (+6.76\% $\uparrow$) & (+3.18\% $\uparrow$) & (+2.36\% $\uparrow$)  \\
    \textbf{~~-Combine}             & 51.14 $\pm$ 0.46 & 28.21 $\pm$ 0.49 & 33.64 $\pm$ 0.56 & 21.13 $\pm$ 0.43 & 25.51 $\pm$ 0.54 & 19.14 $\pm$ 0.55 & 48.29 $\pm$ 0.55 & \textbf{31.56 $\pm$ 0.21} & 95.91 $\pm$ 0.33 & 95.83 $\pm$ 0.32 \\
    \textbf{~~-Time-invariant}         & \textbf{54.02 $\pm$ 0.35} & \textbf{29.98 $\pm$ 0.72} & \textbf{35.40 $\pm$ 0.20} & \textbf{22.17 $\pm$ 0.21} & \textbf{25.69 $\pm$ 0.26} & \textbf{19.80 $\pm$ 0.41} & \textbf{50.55 $\pm$ 0.50} & 31.42 $\pm$ 0.98 & 74.05 $\pm$ 0.99 & 73.32 $\pm$ 1.02 \\
    \textbf{~~-Time-varying}           & 52.45 $\pm$ 0.18 & 23.82 $\pm$ 0.21 & 34.04 $\pm$ 0.55 & 15.12 $\pm$ 0.47 & 25.39 $\pm$ 0.18 & 15.02 $\pm$ 0.31 & 49.03 $\pm$ 0.35 & 24.98 $\pm$ 1.09 & \textbf{96.68 $\pm$ 0.21} & \textbf{95.86 $\pm$ 0.20} \\
    \midrule
    \specialrule{0em}{0.5pt}{0.5pt}
    \midrule
    \textbf{DySAT}                     & 46.65 $\pm$ 3.12 & 25.34 $\pm$ 0.97 & 28.92 $\pm$ 3.18 & 13.54 $\pm$ 1.28 & 25.50 $\pm$ 0.58 & 13.06 $\pm$ 0.46 & 42.95 $\pm$ 2.46 & 25.03 $\pm$ 0.74 & 72.73 $\pm$ 2.06 & 71.21 $\pm$ 2.18\\
    \textbf{DySAT-DyTed}               & (+13.48$\%$ $\uparrow$) & (+10.22$\%$ $\uparrow$) & (+15.08$\%$ $\uparrow$) & (+38.63$\%$ $\uparrow$) & (+9.73$\%$ $\uparrow$) & (+26.26$\%$ $\uparrow$) & (+15.32$\%$ $\uparrow$) & (+14.74$\%$ $\uparrow$) & (+16.72$\%$ $\uparrow$) & (+19.15$\%$ $\uparrow$)           \\
    \textbf{~~-Combine}                 & 49.39 $\pm$ 3.31 & 27.82 $\pm$ 1.76 & 28.07 $\pm$ 4.55 & 13.93 $\pm$ 2.04 & 25.88 $\pm$ 0.52 & 13.12 $\pm$ 0.62 & 45.66 $\pm$ 4.65 & \textbf{33.13 $\pm$ 3.09} & 80.11 $\pm$ 1.96 & 79.98 $\pm$ 2.05\\
    \textbf{~~-Time-invariant}         & \textbf{52.94 $\pm$ 0.48} & \textbf{27.93 $\pm$ 0.50} & \textbf{33.28 $\pm$ 0.73} & \textbf{18.77 $\pm$ 0.94} & \textbf{27.98 $\pm$ 0.71} & \textbf{16.49 $\pm$ 0.82} & \textbf{49.53 $\pm$ 0.42} & 28.72 $\pm$ 1.32 & 71.48 $\pm$ 2.05 & 71.11 $\pm$ 2.07 \\
    \textbf{~~-Time-varying}           & 40.41 $\pm$ 4.87 & 25.16 $\pm$ 1.96 & 23.87 $\pm$ 1.42 & 11.74 $\pm$ 1.08 & 23.06 $\pm$ 1.22 & 11.25 $\pm$ 0.87 & 42.01 $\pm$ 3.10 & 26.05 $\pm$ 0.88 & \textbf{84.89 $\pm$ 1.66} & \textbf{84.85 $\pm$ 1.66} \\
    \midrule
    \specialrule{0em}{0.5pt}{0.5pt}
    \midrule
    \textbf{EvolveGCN}                 & 48.26 $\pm$ 1.01 & 23.89 $\pm$ 1.07 & 31.02 $\pm$ 0.88 & 12.64 $\pm$ 0.74 & 24.39 $\pm$ 1.02 & 11.62 $\pm$ 0.17 & 43.76 $\pm$ 1.00 & 25.76 $\pm$ 0.40 & 92.00 $\pm$ 0.39 & 92.14 $\pm$ 0.39\\
    \textbf{EvolveGCN-DyTed}           & (+11.33$\%$ $\uparrow$) & (+3.47$\%$ $\uparrow$) & (+15.73$\%$ $\uparrow$) & (+19.86$\%$ $\uparrow$) & (+11.19$\%$ $\uparrow$) & (+30.12$\%$ $\uparrow$) & (+15.68$\%$ $\uparrow$) & (+1.28$\%$ $\uparrow$) & (+4.62$\%$ $\uparrow$) & (+5.18$\%$ $\uparrow$)             \\
    \textbf{~~-Combine}           & 53.00 $\pm$ 0.84 & \textbf{24.72 $\pm$ 1.19} & 35.06 $\pm$ 0.64 & 13.43 $\pm$ 0.54 & 26.86 $\pm$ 0.35 & 12.39 $\pm$ 0.18 & 47.96 $\pm$ 1.50 & 26.08 $\pm$ 0.29 & \textbf{96.25 $\pm$ 0.08} & 96.20 $\pm$ 0.06 \\
    \textbf{~~-Time-invariant}         & \textbf{53.73 $\pm$ 0.40} & 23.38 $\pm$ 0.08 & \textbf{35.90 $\pm$ 0.31} & \textbf{15.15 $\pm$ 0.45} & \textbf{27.12 $\pm$ 0.30} & \textbf{15.12 $\pm$ 0.64} & \textbf{50.62 $\pm$ 0.58} & \textbf{26.09 $\pm$ 0.92} & 76.45 $\pm$ 2.11 & 79.71 $\pm$ 2.11 \\
    \textbf{~~-Time-varying}           & 52.53 $\pm$ 1.03 & 24.63 $\pm$ 1.12 & 34.78 $\pm$ 0.58 & 12.11 $\pm$ 0.61 & 26.37 $\pm$ 0.89 & 10.87 $\pm$ 0.17 & 47.82 $\pm$ 1.65 & 24.05 $\pm$ 0.55 & 96.02 $\pm$ 0.08 & \textbf{96.91 $\pm$ 0.08} \\
    \midrule
    \specialrule{0em}{0.5pt}{0.5pt}
    \midrule
    \textbf{HTGN}                      & 54.11 $\pm$ 0.46 & 23.41 $\pm$ 0.13 & 35.74 $\pm$ 0.36 & 10.53 $\pm$ 0.08 & 26.70 $\pm$ 0.39 & 11.25 $\pm$ 0.71 & 50.13 $\pm$ 0.19 & 22.26 $\pm$ 0.06 & 92.25 $\pm$ 1.54 & 92.14 $\pm$ 1.57\\
    \textbf{HTGN-DyTed}                 & (+1.81$\%$ $\uparrow$) & (+1.11$\%$ $\uparrow$) & (+3.47$\%$ $\uparrow$) & (+9.78$\%$ $\uparrow$) & (+6.55$\%$ $\uparrow$) & (+29.69$\%$ $\uparrow$) & (+3.97$\%$ $\uparrow$) & (+4.49$\%$ $\uparrow$) & (+4.80$\%$ $\uparrow$) & (+4.87$\%$ $\uparrow$)             \\
    \textbf{~~-Combine}                & 53.26 $\pm$ 0.55 & 23.17 $\pm$ 0.16 & 36.50 $\pm$ 0.39 & 10.50 $\pm$ 0.08 & 27.43 $\pm$ 0.38 & 12.96 $\pm$ 0.77 & 49.86 $\pm$ 0.36 & 22.18 $\pm$ 0.11 & 96.57 $\pm$ 0.09 & 96.52 $\pm$ 0.09 \\
    \textbf{~~-Time-invariant}         & \textbf{55.09 $\pm$ 0.46} & \textbf{23.67 $\pm$ 0.13} & \textbf{36.98 $\pm$ 0.34} & \textbf{11.56 $\pm$ 0.07} & \textbf{28.45 $\pm$ 0.29} & \textbf{14.59 $\pm$ 0.07} & \textbf{52.12 $\pm$ 0.25} & \textbf{23.26 $\pm$ 0.08} & 93.30 $\pm$ 0.29 & 93.21 $\pm$ 0.29 \\
    \textbf{~~-Time-varying}           & 53.06 $\pm$ 0.37 & 23.11 $\pm$ 0.10 & 35.66 $\pm$ 0.46 & 10.51 $\pm$ 0.10 & 26.99 $\pm$ 0.37 &  8.97 $\pm$ 0.21 & 49.08 $\pm$ 0.30 & 21.95 $\pm$ 0.09 & \textbf{96.68 $\pm$ 0.11} & \textbf{96.63 $\pm$ 0.12} \\
    \midrule
    \specialrule{0em}{0.5pt}{0.5pt}
    \midrule
    \textbf{ROLAND}                    & 50.62 $\pm$ 0.40 & 20.27 $\pm$ 0.11 & 31.21 $\pm$ 0.34 & 10.41 $\pm$ 0.08 & 28.08 $\pm$ 0.43 &  8.77 $\pm$ 0.10 & 49.34 $\pm$ 0.42 & 22.02 $\pm$ 0.12 & 83.27 $\pm$ 0.16 & 75.18 $\pm$ 0.07 \\
    \textbf{ROLAND-DyTed}              & (+9.40$\%$ $\uparrow$) & (+24.32$\%$ $\uparrow$) & (+15.54$\%$ $\uparrow$) & (+20.94$\%$ $\uparrow$) & (+3.74$\%$ $\uparrow$) & (+11.97$\%$ $\uparrow$) & (+5.78$\%$ $\uparrow$) & (+11.58$\%$ $\uparrow$) & (+8.47$\%$ $\uparrow$) & (+7.37$\%$ $\uparrow$)            \\
    \textbf{~~-Combine}              & 54.01 $\pm$ 0.51 & 23.38 $\pm$ 0.14 & 36.04 $\pm$ 0.23 & 11.60 $\pm$ 0.05 & 28.50 $\pm$ 0.32 &  8.87 $\pm$ 0.08 & 50.11 $\pm$ 0.24 & 22.25 $\pm$ 0.07 & 85.45 $\pm$ 0.21 & 75.67 $\pm$ 0.09 \\
    \textbf{~~-Time-invariant}         & \textbf{55.38 $\pm$ 0.36} & \textbf{25.20 $\pm$ 0.10} & \textbf{36.06 $\pm$ 0.47} & \textbf{12.59 $\pm$ 0.10} & \textbf{29.13 $\pm$ 0.37} &  \textbf{9.82 $\pm$ 0.17} & \textbf{52.19 $\pm$ 0.36} & \textbf{24.57 $\pm$ 0.11} & 84.25 $\pm$ 0.20 & 75.17 $\pm$ 0.09 \\
    \textbf{~~-Time-varying}           & 53.76 $\pm$ 0.32 & 23.01 $\pm$ 0.09 & 35.13 $\pm$ 0.34 & 10.40 $\pm$ 0.08 & 27.63 $\pm$ 0.47 &  8.67 $\pm$ 0.12 & 49.15 $\pm$ 0.46 & 21.97 $\pm$ 0.14 & \textbf{90.32 $\pm$ 0.27} & \textbf{80.72 $\pm$ 0.12} \\
    \bottomrule
   
    \end{tabular}
    \end{threeparttable}
}

\label{tab:n-classify}%
\end{table*}%

In this section, we introduce the details of experimental setup.

\subsubsection{\textbf{Datasets}}
In order to evaluate our proposed method, we adopt five commonly used datasets of dynamic graphs, including the communication network UCI~\cite{panzarasa2009patterns}, bitcoin transaction network Bitcoin~\cite{kumar2018rev2}, routing network AS733~\cite{leskovec2005graphs}, citation network HepTh and HepPh~\cite{leskovec2005graphs}. In addition, we also include two real financial datasets of dynamic graphs, Tecent-alpha and Tencent-beta, with high-quality user labels and various downstream tasks. The detailed statistics of all datasets are shown in Table~\ref{tab:datasets}.
\begin{itemize}[leftmargin=*]
    \item \textbf{UCI}~\cite{panzarasa2009patterns} is a communication network, where links represent messages sent between users on an online social network.
    \item \textbf{Bitcoin}~\cite{kumar2018rev2} is a who-trusts-whom network of people who trade using Bitcoin.
    \item \textbf{AS733}~\cite{leskovec2005graphs} is a routing network. The dataset contains a total of 733 snapshots. We follow the setting in~\cite{yang2021discrete} and extract 30 snapshots. To enable all baselines to conduct experiments, we restricted the dataset not to include deleted nodes.
    \item \textbf{HepTh}~\cite{leskovec2005graphs} is a citation network related to high-energy physics theory. 
    We extract 92 months of data from this dataset, forming 23 snapshots. For each reference edge, we set it to exist since the occurrence of the reference.
    \item \textbf{HepPh}~\cite{leskovec2005graphs} is a citation network related to high-energy physics phenomenology. We extract 60 months of data from this dataset, forming 20 snapshots. Other settings are similar to HepTh.
    \item \textbf{Tencent-alpha} is real dynamic graphs of daily capital transactions between users on Tencent after eliminating sensitive information. Each node represents a customer or a business, and each edge represents the transaction. The data ranges from April 1, 2020, to April 30, 2020, with 30 snapshots. Each node has five labels that reflect the time-invariant or time-varying characteristics of users, i.e., annual income grade with five classes, age with five classes, asset with five classes, financing risk with three classes, and consumption fluctuation, i.e. the increase or decrease in daily consumption of users, with binary class.
    \item \textbf{Tencent-beta} has the same data source as Tencent-alpha, but Tencent-beta has more nodes with only two labels, i.e. asset with five classes and financing risk with five classes.
\end{itemize}

\subsubsection{\textbf{Baselines}}
We apply the proposed framework DyTed to the following five baselines. First, we select the combination of GCN~\cite{kipf2016semi} and LSTM (denoted as \textbf{LSTMGCN}) as a basic baseline.
Then we choose three state-of-the-art discrete-time approaches with contrastive or prediction pretext task, i.e., \textbf{DySAT}~\cite{sankar2020dysat} with the contrastive task, \textbf{EvolveGCN}~\cite{pareja2020evolvegcn} and \textbf{HTGN}~\cite{yang2021discrete} with the predictive task.
In addition, we choose a framework \textbf{ROLAND}~\cite{10.1145/3534678.3539300} that extends the static model to the discrete-time dynamic graph representation learning. 
Note that, all the baselines learn representations in a single mixed space.
\begin{itemize}[leftmargin=*]
    \item \textbf{DySAT}~\cite{sankar2020dysat} computes node representations through joint self-attention along the two dimensions of the structural neighborhood and temporal dynamics.
    \item \textbf{EvolveGCN}~\cite{pareja2020evolvegcn} adapts the GCN to compute node representations, and captures the dynamism of the graph sequence by using an RNN to evolve the GCN parameters.
    \item \textbf{HTGN}~\cite{yang2021discrete} maps the dynamic graph into hyperbolic space, and incorporates hyperbolic graph neural network and hyperbolic gated recurrent neural network to obtain representations.
    \item \textbf{ROLAND}~\cite{10.1145/3534678.3539300} views the node representations at different GNN layers as hierarchical node states and recurrently updates them.
\end{itemize}

\subsubsection{\textbf{Implementation Details}}
For all baselines and our DyTed, the representation model is trained on the snapshots $\{\mathcal{G}^1,\mathcal{G}^2, ..., \mathcal{G}^T \}$. 
The total dimension of the representation is set according to the scale of datasets, chosen from $\{16, 32, 64, 128\}$.
We ensure that the dimensions of the representation of each method are equal on the same dataset. We follow the hyperparameters given in the paper of the baselines. The stop of the training process depends on the loss of the pretext task.
The learning rate is chosen from $\{0.1, 0.01, ..., 10^{-5}\}$. In our framework, we choose $\lambda_1$ and $\lambda_2$ from $\{0.1, 0.2,..., 1.0\}$, and set $\lambda_3$ to $5e-7$, temperature parameter $\tau$ to $0.1$ and $\tau_g$ to $0.01$. ${\rm sim}(\cdot)$ mentioned in Section~\ref{sec:method} is a cosine function.
Our implementation code is provided through the link~\footnote{https://github.com/Kaike-Zhang/DyTed}.

\subsection{Performance on Downstream Tasks}
\label{sec:performance}

In this section, we answer RQ1, i.e., whether the DyTed framework can improve the performance of existing methods in various downstream tasks?
We evaluate the performance of our method using three diverse tasks: node classification with time-invariant labels, node classification with the time-varying label, and link prediction for the next snapshot.
For clarity, we use some abbreviations to denote the different representations as follows:
\begin{itemize}[leftmargin=*]
    \item \textbf{Baseline}: the representations from the original baselines, , i.e., $\bm{r}_v^t \in \mathbb{R}^d, t = 1,2,...,T$ for node $v$.
    \item \textbf{Baseline-DyTed}: apply the proposed DyTed framework to the baseline method.
    \item \textbf{~~-Combine}: the combination of time-invariant and time-varying representation, i.e., $\bm{r}_v^t = (\bm{s}_v, \bm{d}_v^t) \in \mathbb{R}^d, t = 1,2,...,T$ for node $v$.
    \item \textbf{~~-Time-invariant}: only the time-invariant representation $\bm{s}_v \in \mathbb{R}^{\frac{d}{2}} $ for node $v$.
    \item \textbf{~~-Time-varying}: only the time-varying representation $\bm{d}_v^t \in \mathbb{R}^{\frac{d}{2}} , t = 1,2,...,T$ for node $v$.
\end{itemize}

\begin{table}
\centering
\caption{Node classification on Tencent-beta}
\resizebox{3.4in}{!}{
    \begin{threeparttable}
    \begin{tabular}{lcccc}
    
    \toprule
    \multirow{2}[2]{*}{\textbf{Model}} & \multicolumn{2}{c}{\textbf{Assets}} & \multicolumn{2}{c}{\textbf{Finacing Risk}}\\
                                         \cmidrule(r){2-3}                            \cmidrule(r){4-5}
                                       & \textbf{micro-F1} & \textbf{macro-F1}      & \textbf{micro-F1} & \textbf{macro-F1}\\
    \midrule
    \textbf{LSTMGCN}                   & 33.03 $\pm$ 0.51 & 18.95 $\pm$ 0.87 & 29.59 $\pm$ 0.52 & 19.25 $\pm$ 0.52\\
    \textbf{LSTMGCN-DyTed}             & (+11.96\% $\uparrow$) & (+6.39\% $\uparrow$) & (+9.33\% $\uparrow$) & (+10.81\% $\uparrow$) \\
    \textbf{~~-Combine}                & 36.61 $\pm$ 0.45 & 20.08 $\pm$ 0.41 & 30.64 $\pm$ 0.60 & 20.10 $\pm$ 0.29 \\
    \textbf{~~-Time-invariant}         & \textbf{36.98 $\pm$ 0.30} & \textbf{20.16 $\pm$ 0.22} & \textbf{32.35 $\pm$ 0.41} & \textbf{21.33 $\pm$ 0.43} \\
    \textbf{~~-Time-varying}           & 34.65 $\pm$ 0.40 & 16.78 $\pm$ 0.59 & 30.36 $\pm$ 0.37 & 19.43 $\pm$ 0.48 \\
    \midrule
    \specialrule{0em}{0.5pt}{0.5pt}
    \midrule
    \textbf{DySAT}                      & OOM \tnote{1} & OOM & OOM & OOM \\
    \midrule
    \specialrule{0em}{0.5pt}{0.5pt}
    \midrule
    \textbf{EvolveGCN}                 & 39.21 $\pm$ 0.40 & 22.17 $\pm$ 0.46 & 37.56 $\pm$ 0.70 & 22.16 $\pm$ 1.33\\
    \textbf{EvolveGCN-DyTed}           & (+5.43\% $\uparrow$) & (+12.00\% $\uparrow$) & (+1.54\% $\uparrow$) & (+14.89\% $\uparrow$) \\
    \textbf{~~-Combine}                 & \textbf{41.34 $\pm$ 0.76} & 24.38 $\pm$ 0.47 & 37.78 $\pm$ 0.71 & 25.19 $\pm$ 0.31 \\
    \textbf{~~-Time-invariant}         & 40.78 $\pm$ 0.46 & \textbf{24.83 $\pm$ 0.29} & \textbf{38.14 $\pm$ 0.72} & \textbf{25.46 $\pm$ 0.29} \\
    \textbf{~~-Time-varying}           & 39.07 $\pm$ 0.37 & 21.63 $\pm$ 0.31 & 37.01 $\pm$ 0.43 & 25.03 $\pm$ 0.67 \\
    \midrule
    \specialrule{0em}{0.5pt}{0.5pt}
    \midrule
    \textbf{HTGN}                       & OOM & OOM & OOM & OOM \\
    \midrule
    \specialrule{0em}{0.5pt}{0.5pt}
    \midrule
    \textbf{ROLAND}                    & 33.59 $\pm$ 0.15 & 10.06 $\pm$ 0.04 & 33.05 $\pm$ 0.37 & 9.93 $\pm$ 0.08 \\
    \textbf{ROLAND-DyTed}              & (+10.87\% $\uparrow$) & (+103.38\% $\uparrow$) & (+6.20\% $\uparrow$) & (+63.44\% $\uparrow$)  \\
    \textbf{~~-Combine}                & 36.01 $\pm$ 0.65 & 19.11 $\pm$ 0.56 & 34.47 $\pm$ 0.74 & 15.61 $\pm$ 0.83 \\
    \textbf{~~-Time-invariant}         & \textbf{37.24 $\pm$ 1.01} & \textbf{20.46 $\pm$ 0.21} & \textbf{35.10 $\pm$ 0.47} & \textbf{16.23 $\pm$ 0.81} \\
    \textbf{~~-Time-varying}           & 33.23 $\pm$ 0.31 & 9.98 $\pm$ 0.07 & 33.24 $\pm$ 0.17 & 9.98 $\pm$ 0.04 \\
    \bottomrule
   
    \end{tabular}
    \begin{tablenotes}
      \item[1] Out of memory.
    \end{tablenotes}
    \end{threeparttable}
}

\label{tab:appendix_ncls}%
\end{table}%

\begin{table*}[t]
\centering
\caption{Link prediction $^1$}
\resizebox{\textwidth}{!}{
    \begin{threeparttable}
    \begin{tabular}{lcccccccccc}
    
    \toprule
    \multirow{2}[2]{*}{\textbf{Model}} & \multicolumn{2}{c}{\textbf{Uci}} & \multicolumn{2}{c}{\textbf{Bitcoin}}        & \multicolumn{2}{c}{\textbf{as733}}    & \multicolumn{2}{c}{\textbf{HepTh}}   & \multicolumn{2}{c}{\textbf{HepPh}} \\
                                         \cmidrule(r){2-3}                            \cmidrule(r){4-5}                         \cmidrule(r){6-7}                        \cmidrule(r){8-9}                               \cmidrule(r){10-11}
                                       & \textbf{AUC} & \textbf{AP}      & \textbf{AUC} & \textbf{AP}   & \textbf{AUC} & \textbf{AP}  & \textbf{AUC} & \textbf{AP}         & \textbf{AUC} & \textbf{AP}\\
    \midrule
    \textbf{LSTMGCN}                   & 68.64 $\pm$ 0.23 & 68.04 $\pm$ 1.28 & 72.91 $\pm$ 0.87 & 72.69 $\pm$ 0.85 & 74.68 $\pm$ 0.57 & 75.11 $\pm$ 0.58 & 78.87 $\pm$ 1.01 & 78.55 $\pm$ 0.96 & 73.34 $\pm$ 1.29 & 73.50 $\pm$ 1.35\\
    \textbf{LSTMGCN-DyTed}             & 76.06 $\pm$ 0.42 & 76.61 $\pm$ 0.59 & 79.02 $\pm$ 0.70 & 78.91 $\pm$ 0.64 & 81.82 $\pm$ 0.89 & 81.10 $\pm$ 1.07 & 85.88 $\pm$ 0.23 & 85.83 $\pm$ 0.27 & 87.12 $\pm$ 0.45 & 86.94 $\pm$ 0.52 \\
    \midrule
    \specialrule{0em}{0.5pt}{0.5pt}
    \midrule
    \textbf{DySAT}                     & 74.99 $\pm$ 0.53 & 79.91 $\pm$ 1.19 & 77.50 $\pm$ 1.46 & 77.45 $\pm$ 1.39 & 80.91 $\pm$ 0.79 & 81.00 $\pm$ 0.82 & 78.88 $\pm$ 0.58 & 79.35 $\pm$ 0.63 & 79.40 $\pm$ 1.30 & 74.98 $\pm$ 1.27\\
    \textbf{DySAT-DyTed}               & 86.12 $\pm$ 0.46 & 85.30 $\pm$ 0.69 & 83.32 $\pm$ 1.03 & 83.40 $\pm$ 1.00 & 84.47 $\pm$ 0.89 & 84.75 $\pm$ 0.90 & 80.76 $\pm$ 0.43 & 81.02 $\pm$ 0.51 & 78.68 $\pm$ 1.56 & 78.64 $\pm$ 1.60 \\
    \midrule
    \specialrule{0em}{0.5pt}{0.5pt}
    \midrule
    \textbf{EvolveGCN}                 & 79.78 $\pm$ 0.50 & 84.01 $\pm$ 1.05 & 78.69 $\pm$ 1.60 & 78.96 $\pm$ 1.69 & 82.99 $\pm$ 1.23 & 83.18 $\pm$ 1.22 & 81.56 $\pm$ 0.74 & 81.28 $\pm$ 0.78 & 77.12 $\pm$ 1.16 & 77.60 $\pm$ 1.19\\
    \textbf{EvolveGCN-DyTed}            & 84.15 $\pm$ 1.25 & 85.88 $\pm$ 1.17 & 85.50 $\pm$ 0.48 & 85.35 $\pm$ 0.54 & 80.97 $\pm$ 1.66 & 83.96 $\pm$ 1.25 & 87.52 $\pm$ 0.23 & 87.48 $\pm$ 0.22 & 80.21 $\pm$ 0.78 & 80.27 $\pm$ 0.78 \\
    \midrule
    \specialrule{0em}{0.5pt}{0.5pt}
    \midrule
    \textbf{HTGN}                      & 88.52 $\pm$ 0.63 & 88.93 $\pm$ 0.76 & 80.23 $\pm$ 1.62 & 80.42 $\pm$ 1.59 & 73.17 $\pm$ 1.33 & 73.16 $\pm$ 1.31 & 70.51 $\pm$ 0.45 & 72.79 $\pm$ 0.63 & 72.77 $\pm$ 0.84 & 72.34 $\pm$ 0.81\\
    \textbf{HTGN-DyTed}                & 92.08 $\pm$ 0.14 & 91.85 $\pm$ 0.37 & 84.56 $\pm$ 0.29 & 84.15 $\pm$ 0.19 & 77.08 $\pm$ 0.11 & 76.94 $\pm$ 0.25 & 75.61 $\pm$ 0.23 & 75.88 $\pm$ 0.21 & 74.46 $\pm$ 0.57 & 74.37 $\pm$ 0.55 \\
    \midrule
    \specialrule{0em}{0.5pt}{0.5pt}
    \midrule
    \textbf{ROLAND}                    & 86.88 $\pm$ 0.96 & 86.73 $\pm$ 0.12 & 85.67 $\pm$ 0.92 & 85.09 $\pm$ 0.90 & 74.86 $\pm$ 0.46 & 73.01 $\pm$ 0.57 & 80.81 $\pm$ 0.15 & 81.43 $\pm$ 0.20 & 80.16 $\pm$ 1.02 & 80.46 $\pm$ 1.35 \\
    \textbf{ROLAND-DyTed}              & 88.05 $\pm$ 0.63 & 89.91 $\pm$ 0.13 & 88.59 $\pm$ 0.06 & 88.63 $\pm$ 0.13 & 76.96 $\pm$ 0.23 & 74.28 $\pm$ 0.35 & 81.14 $\pm$ 0.67 & 79.90 $\pm$ 0.53 & 80.50 $\pm$ 0.26 & 81.00 $\pm$ 0.30 \\
    \bottomrule
    \end{tabular}
    \begin{tablenotes}
         \item[1] See Appendix~\ref{appendix:supp} for the performance on time-invariant representations and time-varying representations in Table~\ref{tab:appendix_link}.
    \end{tablenotes}
    \end{threeparttable}
}

\label{tab:l-prediction}%
\end{table*}%

\subsubsection{\textbf{Node Classification with Time-invariant Labels}}
In this section, we evaluate the performance of each method on node classification with time-invariant labels, i.e., annual income, age, asset, and financing risk. Considering that the time range of graphs is only one month, the above labels can well reflect the identity or stable characteristics of users.
We employ a linear layer with softmax as the downstream classifier and take the representation of the snapshot $\mathcal{G}^{T}$ as the classifier’s input for all baselines. \textbf{Note that}, we have also tried to take the pooling of the representations over all snapshots as a time-invariant representation, but gained poor performance (see Table~\ref{tab:appendix_ncls_avg} in Appendix~\ref{appendix:poolrep}). 
We divide downstream labels according to $0.2:0.2:0.6$ to serve as the train, validation, and test set.
The results are shown in Table~\ref{tab:n-classify} and Table~\ref{tab:appendix_ncls}. It can be observed that when applying our framework to existing methods, the performance is significantly improved, demonstrating the effectiveness of the disentanglement.
Specifically, the time-invariant representation disentangled via our framework achieves the best performance, gaining an average improvement of 8.81\% in micro-F1 and 16.01\% in macro-F1 on Tencent-alpha, and an average improvement of 7.55\% in micro-F1 and 35.15\% in macro-F1 on the large dynamic graph dataset Tencent-beta for such time-invariant classification tasks.
It is worth noting that the dimension of the time-invariant representation is even half that of the baseline representation.

\subsubsection{\textbf{Node Classification with Time-varying Labels}}
In this subsection, we take the node classification with a time-invariant label as the downstream evaluation task. Specifically, we take the consumption fluctuation of users as the label, i.e. the increase or decrease in daily consumption of users.
From Table~\ref{tab:n-classify}~(last two columns), similarly, we can see that the framework DyTed significantly improves the performance of baselines, where the time-varying representation achieves the best performance and gain an average improvement of 7.51\% in micro-F1 and 7.79\% in macro-F1.

\subsubsection{\textbf{Link Prediction}}
We use representations of two nodes at snapshot $\mathcal{G}^{T}$ to predict whether they are connected at the snapshot $\mathcal{G}^{T+1}$, which is a commonly adopted evaluation task for dynamic graph representation learning. We follow the evaluation method used in \cite{sankar2020dysat}: taking logistic regression as the classifier and the Hadamard product of representations as the input.
Table~\ref{tab:l-prediction} shows that our framework consistently improves baselines on all datasets with an average improvement of 5.87\% in AUC and 5.76\% in AP.

\begin{table}[t]
  \centering
  \caption{Ablation study}
  \resizebox{3.4in}{!}{
  \begin{threeparttable}
    \begin{tabular}{lccccc}
    \toprule
    \textbf{Model} & \textbf{Uci} &\textbf{Bitcoin} &\textbf{As722} &\textbf{HepTh} &\textbf{HepPh} \\
    \midrule
    \textbf{LSTMGCN-DyTed}                            & $\bm{76.06}$ & $\bm{79.02}$ & $\bm{81.82}$ & $\bm{85.88}$ & $\bm{87.12}$   \\
    \hspace{0.3cm}-Random Sampling                    & $71.09$ & $77.72$ & $80.57$ & $85.79$ & $87.03$\\
    \hspace{0.3cm}-w/o Time-varying Generator         & $72.85$ & $69.83$ & $73.69$ & $76.49$ & $82.78$\\
    \hspace{0.3cm}-w/o Discriminator                  & $76.03$ & $76.88$ & $80.52$ & $85.58$ & $86.12$  \\
    \midrule
    \specialrule{0em}{0.5pt}{0.5pt}
    \midrule
    \textbf{DySAT-DyTed}                              & $\bm{86.12}$ & $\bm{83.32}$ & $\bm{84.47}$ & $\bm{80.76}$ & $\bm{78.68}$   \\
    \hspace{0.3cm}-Random Sampling                    & $74.24$ & $82.91$ & $83.04$ & $79.26$ & $75.69$ \\
    \hspace{0.3cm}-w/o Time-varying Generator         & $71.26$ & $79.61$ & $78.7$ & $74.87$ & $71.66$ \\
    \hspace{0.3cm}-w/o Discriminator                  & $78.20$ & $80.21$ & $84.38$ & $77.47$ & $76.01$  \\
    \midrule
    \textbf{EvolveGCN-DyTed}                          & $\bm{84.15}$ & $\bm{85.50}$ & $\bm{80.97}$ & $\bm{87.52}$ & $\bm{80.21}$  \\
    \hspace{0.3cm}-Random Sampling                    & $80.62$ & $79.14$ & $79.85$ & $78.22$ & $79.64$\\
    \hspace{0.3cm}-w/o Time-varying Generator         & $82.55$ & $78.85$ & $78.10$ & $67.12$ & $73.19$\\
    \hspace{0.3cm}-w/o Discriminator                  & $83.75$ & $84.15$ & $77.65$ & $79.79$ & $70.54$ \\
    \midrule
    \specialrule{0em}{0.5pt}{0.5pt}
    \midrule
    \textbf{HTGN-DyTed}                               & $\bm{92.08}$ & $\bm{84.56}$ & $\bm{77.08}$ & $\bm{75.61}$ & $\bm{74.46}$  \\
    \hspace{0.3cm}-Random Sampling                    & $85.09$ & $82.21$ & $76.00$ & $73.05$ & $74.26$\\
    \hspace{0.3cm}-w/o Time-varying Generator         & $85.73$ & $70.54$ & $76.92$ & $73.81$ & $73.85$\\
    \hspace{0.3cm}-w/o Discriminator                  & $90.93$ & $84.01$ & $77.01$ & $72.31$ & $74.21$ \\
    \midrule
    \specialrule{0em}{0.5pt}{0.5pt}
    \midrule
    \textbf{ROLAND-DyTed}                            & $\bm{88.05}$ & $\bm{88.59}$ & $\bm{76.96}$ & $\bm{81.14}$ & $\bm{80.50}$\\
    \hspace{0.3cm}-Random Sampling                   & $84.11$ & $86.32$ & $74.89$ & $78.11$ & $79.19$ \\
    \hspace{0.3cm}-w/o Time-varying Generator        & $73.97$ & $86.95$ & $74.83$ & $78.72$ & $79.87$\\
    \hspace{0.3cm}-w/o Discriminator                 & $85.76$ & $84.34$ & $74.16$ & $80.47$ & $80.10$ \\
    \bottomrule
    \end{tabular}%
    \begin{tablenotes}
         \item[1] The DyTed without the time-invariant representation generator is equivalent to the original backbone model.
    \end{tablenotes}
    \end{threeparttable}
    }
  \label{tab:ab-study}%
\end{table}%

\subsection{Analysis of Disentangled Representation}
In this section, we answer RQ2, i.e., what does each component of DyTed bring? We conduct the ablation study and illustrate the effectiveness of the disentanglement-aware discriminator.

\subsubsection{\textbf{Ablation Study}}
We name different versions as follows:
\begin{itemize}[leftmargin=*]
    \item \textbf{-DyTed-Radom Sampling}: Replace bidirectional Bernoulli sampling with random sampling.
    \item \textbf{-DyTed-w/o-Time-varying Generator}: Remove the time-varying representation generator.
    \item \textbf{-DyTed-w/o-Discriminator}: Remove the disentanglement-aware discriminator (adversarial learning).
\end{itemize}

As shown in Table~\ref{tab:ab-study}, the incorporation of bidirectional Bernoulli sampling significantly improves the model performance when compared with random sampling, demonstrating the effectiveness of our designed sampling strategy in finding cost-effective pairs of temporal clips. Additionally, we find it not easy to obtain satisfactory results using only a time-invariant generator as a standalone method, indicating that modeling solely for time-invariant properties is insufficient. The introduction of the discriminator also contributes to the performance, showing that the better the representation disentangle, the more useful it is for downstream tasks.

\begin{figure}
    \centering
    \includegraphics[width=3.2in]{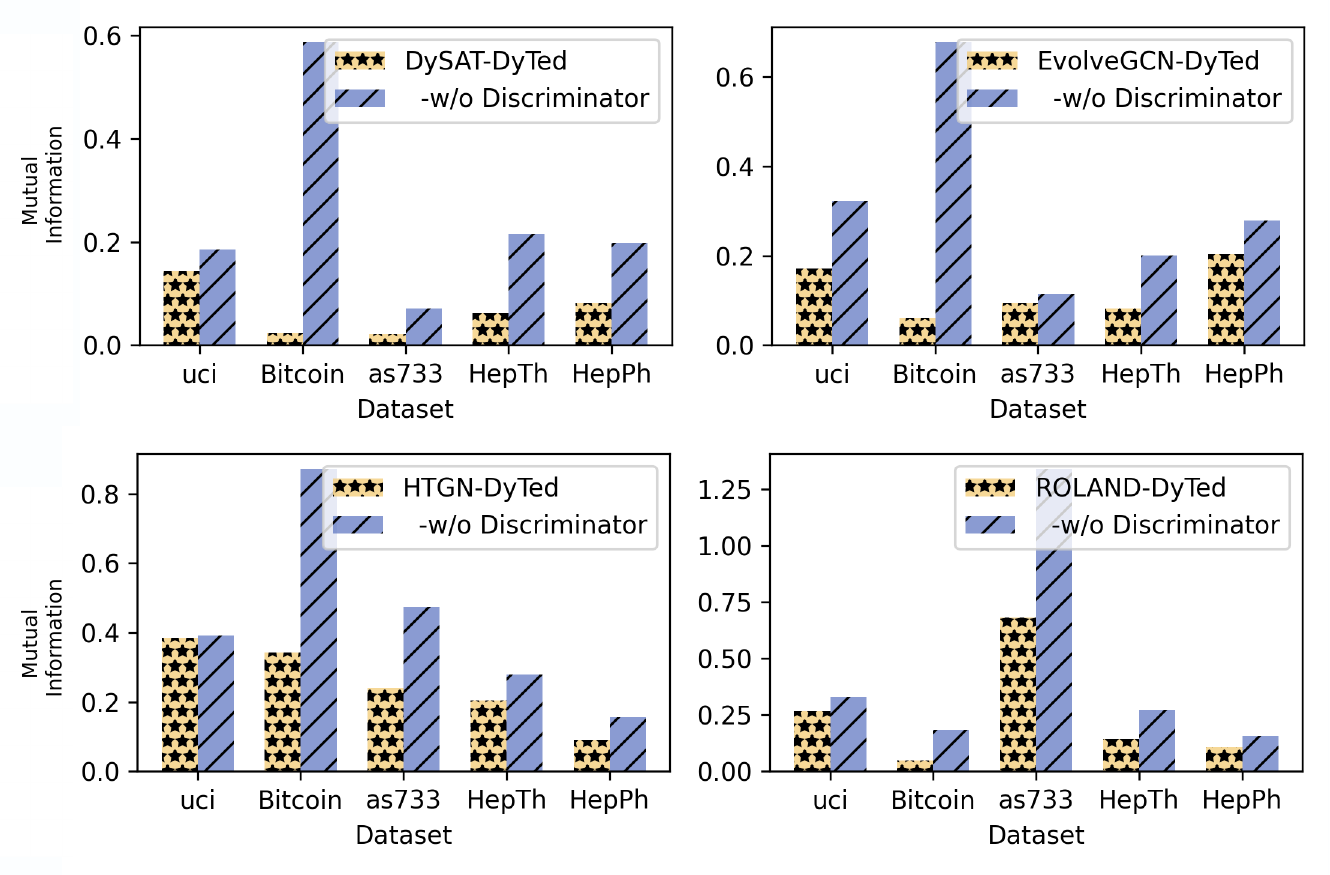} 
    \caption{ The discriminator can effectively minimize the mutual information between time-invariant representations and time-varying representations.
    }
    \label{fig:mutual_info}
\end{figure}

\begin{figure}
    \centering
    \includegraphics[width=3.2in]{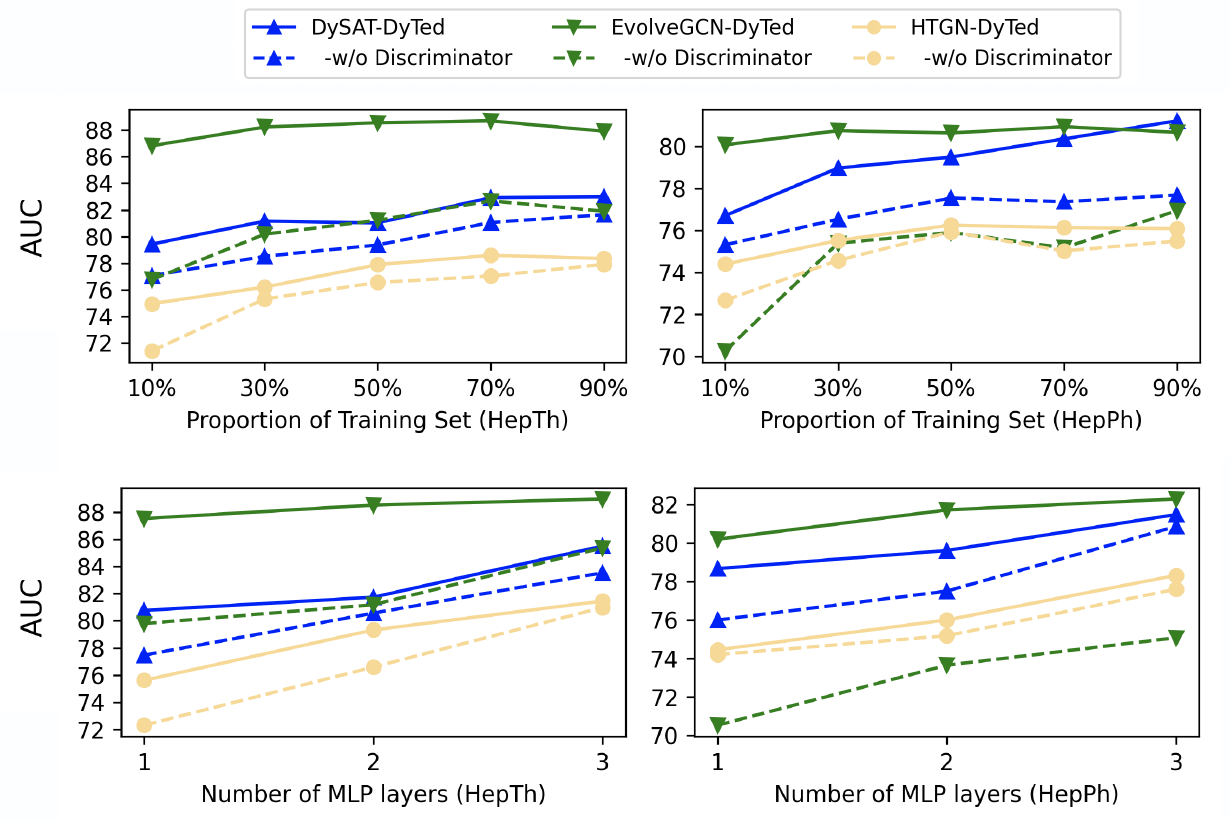}
    \caption{
        The performance with different training resources (training data and model complexity) for downstream tasks.
    }
    \label{fig:trainig_res}
\end{figure}

\subsubsection{\textbf{Evaluation of Disentangling Degree}}
To further quantitatively evaluate the effectiveness of the discriminator in disentangling the time-invariant and the time-varying representations, we measure the mutual information between these two types of representations. 
As illustrated in Figure~\ref{fig:mutual_info}, the designed discriminator is able to further reduce the mutual information between the time-invariant and the time-varying representations, resulting in a higher degree of disentanglement.

\subsection{Benefit of the Disentanglement}
\label{sec:benefit}
In this section, we answer RQ3, i.e., is there any additional benefit of disentangling representation? In particular, we analyze such benefits in terms of both training resources required by downstream tasks and the robustness of models. Here we only show part of the results for the readability of figures, more results see Figure~\ref{fig:appresources2}, Figure~\ref{fig:approbustness2}, Figure~\ref{fig:appresources}, and Figure~\ref{fig:approbustness} in Appendix~\ref{appendix:supp}.

\subsubsection{\textbf{Training Resources}}
As demonstrated in Figure~\ref{fig:trainig_res}, the representation disentangled through the discriminator is able to achieve good performance with a smaller proportion of training data or a simpler downstream model. In contrast, the framework without discriminator needs more training resources to obtain a comparable or even lower performance.

\subsubsection{\textbf{Robustness against noise}}
\begin{figure}
    \centering
    \includegraphics[width=3.2in]{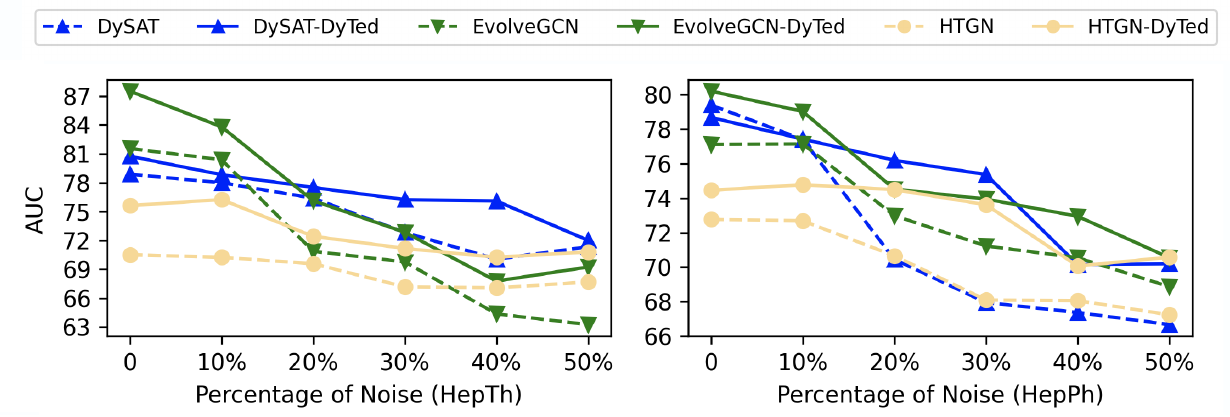}
    \caption{
        Performance against different noise rates.
    }
    \label{fig:robustness}
\end{figure}

To evaluate the robustness of our framework DyTed, we add noise to the original data by randomly adding or deleting edges for each snapshot. The ratio $r\%$ of noise refers to the proportion of added or deleted edges to existing edges, which increased from 0\% to 50\% in steps of 10\%.
As shown in Figure~\ref{fig:robustness}, our framework achieves better robustness, i.e., when compared with baselines, the performance of DyTed decreases less with the percentage of noise increases.

\subsection{Hyper-parameter Analysis}

\begin{figure}
    \centering
    \includegraphics[width=3.2in]{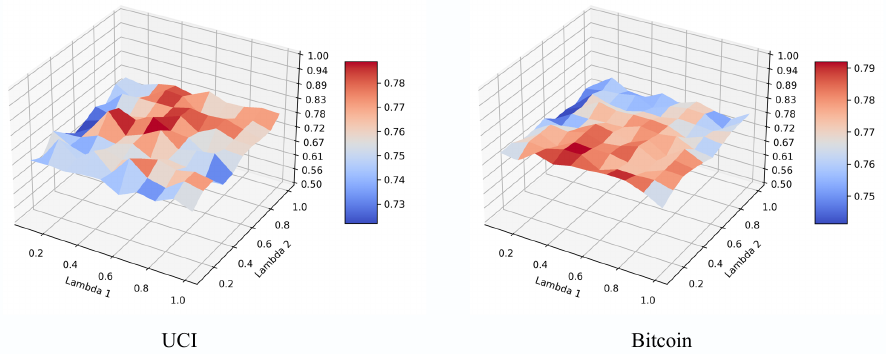}
    \caption{
        Hyper-parameter analysis of $\lambda_1$ and $\lambda_2$.
    }
    \label{fig:hyper_parameter}
\end{figure}

We also analyze the effect of the hyper-parameters $\lambda_1$ and $\lambda_2$ on the performance of our proposed framework DyTed.
We choose $\lambda_1$ and $\lambda_2$ from $\{0.1,0.2,...,1.0\}$, and take the LSTMGCN as an exampled backbone model.
As illustrated in Figure~\ref{fig:hyper_parameter}, the overall performance of the framework is relatively stable. 

\section{CONCLUSION}
In this paper, we propose a novel disentangled representation learning framework for discrete-time dynamic graphs, namely DyTed. 
We propose a time-invariant representation generator with temporal clips contrastive learning, together with a time-varying generator with predefined pretext tasks. To improve the optimization efficiency, we design the bidirectional Bernoulli sampling to higher the sampling probability of cost-effective pairs in contrastive learning. Moreover, to further enhance the disentanglement of the two types of representations, we propose a disentanglement-aware discriminator under an adversarial learning framework from the perspective of information theory.
Extensive experiments demonstrate the effectiveness, robustness, and generality of our framework.

In this work, due to the limitations of data, we only consider discrete-time dynamic graph representation learning methods. In the future, we devote ourselves to extending our framework to more types of dynamic graph representation learning approaches and scenarios, such as continuous-time dynamic graph methods.

\section{Acknowledgments}
This work is funded by the National Key R\&D Program of China (2022YFB3103700, 2022YFB3103701), and the National Natural Science Foundation of China under Grant Nos. U21B2046, 62272125, 62102402. This work is also sponsored by CCF-Tencent Open Fund (RAGR20210108). Huawei Shen is also supported by Beijing Academy of Artificial Intelligence (BAAI).

\bibliographystyle{ACM-Reference-Format}
\bibliography{ref}

\clearpage
\appendix
\section{APPENDIX}
\subsection{Temporal-clips Sampling}
\label{appendix:temporal-clips_samling}

We conduct the following experiments to get some intuitions for effective pairs of temporal clips for optimization. Specifically, according to whether the sampled pair of temporal clips have overlapping snapshots, we divide all pairs into two types: overlapping pairs and non-overlapping pairs. We then optimize one type of pairs with a given length of temporal clips and observe the loss changes on all types of pairs.

As shown in Figure~\ref{fig:smp}, we have three interesting observations: 1). We find that, on average, optimizing non-overlapping pairs benefits more for the overall loss reduction than overlapping pairs, i.e., the yellow dotted line is lower than the blue dotted line. This phenomenon is intuitive since the pair of two overlapping temporal clips is naturally more likely to get similar representations, which may thus have less effect on the optimization of other pairs. 2). For non-overlapping pairs, optimizing long pairs are more effective than short pairs. This may be because long temporal clips cover more snapshots, forming more difficult samples for optimization. 3). For overlapping pairs, the result is reverse, that is, optimizing long pairs are less effective. This may be due to the fact that the number of overlapped snapshots will also be more in the long overlapping pair of temporal clips, resulting in less effective optimization.

\subsection{Proof for Proposition~\ref{pro:overlap}}
\label{proof:overlap}
\begin{proof} 
    First, give 
    \begin{footnotesize}
    $Pr(M=m|L=l)=\Phi(l)^{-1}p(1-p)^{(m-1)}$
    \end{footnotesize}
    , we have the probability 
    \begin{footnotesize}
    $Pr(X|L)$ 
    \end{footnotesize}
    as follows:
    \begin{footnotesize}
        \setlength{\jot}{5pt}
        \begin{equation}
        \label{eq:ap_px}
        \begin{aligned}
            Pr(X=1|L=l) &= \sum_{m_1=1}^lPr(M=m_1|L=l)\sum_{m_2=l-m_1+1}^lPr(M=m_2|L=l)\\
            &= \Phi(l)^{-2}[p \cdot l(1-p)^{l-1}-(1-p)^l+(1-p)^{2l}],\\
            Pr(X=0|L=l) &= \sum_{m_1=1}^lPr(M=m_1|L=l)\sum_{m_2=1}^{l-m_1}Pr(M=m_2|L=l)\\
            &= \Phi(l)^{-2}[1-(1-p)^l-p\cdot l(1-p)^{l-1}],\\
        \end{aligned}
        \end{equation}
    \end{footnotesize}
    where 
    \begin{footnotesize}
    $\Phi(l) = \sum_{m=1}^l p(1-p)^{(m-1)}$.
    \end{footnotesize}
    
    Since  
    \begin{footnotesize}
    $Pr(L=l) = \dfrac{1}{T}$
    \end{footnotesize}
    , we have:
    \begin{footnotesize}
    \setlength{\jot}{5pt}
    \begin{equation}
    \begin{aligned}
         Pr(X=1) - Pr(X=0) =& \sum_{l=0}^TPr(L=l)\left(Pr\left(X=1|L=l\right)-Pr\left(X=0|L=l\right)\right) \\
         =& \dfrac{1}{T} \sum_{l=0}^T \Phi(l)^{-2}(2pl(1-p)^{l-1}+(1-p)^{2l}-1) \\
    \end{aligned}
    \end{equation}
    \end{footnotesize}
    
    Let 
    \begin{footnotesize}
    $R(p,l) = 2pl(1-p)^{l-1}+(1-p)^{2l}-1$
    \end{footnotesize}
    , when 
    \begin{footnotesize}
    $l \geq 3$ 
    \end{footnotesize}
    and 
    \begin{footnotesize}
    $\dfrac{2}{l+2} \leq p < 1$
    \end{footnotesize}
    , we have 
    \begin{footnotesize}
    $\nabla_p R(p,l) \leq 0$
    \end{footnotesize}
    , and 
    \begin{footnotesize}
    $\nabla_l R(p,l) \leq 0$.
    \end{footnotesize}
    Then:
    \begin{footnotesize}
    \begin{equation}
    \begin{aligned}
         Pr(X=1) - Pr(X=0) \leq \dfrac{1}{T} \sum \Phi(l)^{-2}(R(p=\dfrac{2}{5}, l=3)) \leq 0
    \end{aligned}
    \end{equation}
    \end{footnotesize}

    In other words, we have: 
    \begin{footnotesize}
    \begin{equation}
    \label{eq:ap_p10}
    \begin{aligned}
            \dfrac{Pr(X=1)}{Pr(X=0)} \leq 1
    \end{aligned}
    \end{equation}
    \end{footnotesize}
    Accroding to Eq.~\ref{eq:ap_px}:
    \begin{footnotesize}
    \setlength{\jot}{5pt}
    \begin{equation}
    \label{eq:proof_1}
    \begin{aligned} 
         \dfrac{Pr(L=l+1|X=1)}{Pr(L=l|X=1)} = & \dfrac{\Phi(l)^2Pr(L=l+1)}{\Phi(l+1)^2Pr(L=l)}\\
           & \cdot \dfrac{p\cdot (l+1)(1-p)^{l}-(1-p)^{l+1}+(1-p)^{2l+2}}{p\cdot l(1-p)^{l-1}-(1-p)^l+(1-p)^{2l}}  
    \end{aligned}
    \end{equation}
    \end{footnotesize}
    
    Since 
    \begin{footnotesize}
    $0 < \dfrac{\Phi(l)^2}{\Phi(l+1)^2} \leq 1$ 
    \end{footnotesize}
    , and
    \begin{footnotesize}
    $\dfrac{Pr(L=l+1|X=1)}{Pr(L=l|X=1)}\geq 0$
    \end{footnotesize}
    , the Eq.~\ref{eq:proof_1} can be written as: 
    \begin{footnotesize}
    \setlength{\jot}{5pt}
    \begin{equation}
    \begin{aligned}
         \dfrac{Pr(L=l+1|X=1)}{Pr(L=l|X=1)} \leq & \dfrac{p \cdot (l+1)(1-p)^{l}-(1-p)^{l+1}+(1-p)^{2l+2}}{p\cdot l(1-p)^{l-1}-(1-p)^l+(1-p)^{2l}} \\
         =& \dfrac{p \cdot l(1-p)^{l}-(1-p)^{l+1}+(1-p)^{2l+2}+p(1-p)^l}{p \cdot l(1-p)^{l-1}-(1-p)^l+(1-p)^{2l}}\\
         \leq & (1-p) + \dfrac{p(1-p)^l}{p \cdot l(1-p)^{l-1}-(1-p)^l+(1-p)^{2l}}\\
         \leq & (1-p) + \dfrac{p(1-p)^l}{p \cdot l(1-p)^{l-1}-(1-p)^l}\\
         =  & 1 + \dfrac{p\left(2-p\left(l+2\right)\right)}{p\cdot \left(l+1\right) - 1}
    \end{aligned}
    \end{equation}
    \end{footnotesize}

    Given that 
    \begin{footnotesize}
    $\dfrac{2}{l+2}\leq p <1$
    \end{footnotesize}
    , we have 
    \begin{footnotesize}
    $\left(2-p\left(l+2\right)\right) \leq 0$ 
    \end{footnotesize}
    and 
    \begin{footnotesize}
    $p\cdot \left(l+1\right) - 1 >0$
    \end{footnotesize}
    . Then we have:
    
    \begin{footnotesize}
    \begin{equation}
    \label{eq:ap_p1}
    \begin{aligned}
        \dfrac{Pr(L=l+1|X=1)}{Pr(L=l|X=1)}= \dfrac{Pr(X=1|L=l+1)}{Pr(X=1|L=l)} \leq 1 
    \end{aligned}
    \end{equation}
    \end{footnotesize}

    Since $Pr(X=1|L=l) + Pr(X=0|L=l) = 1$, we then have:
    \begin{footnotesize}
    \setlength{\jot}{5pt}
    \begin{equation}
    \label{eq:ap_p0}
    \begin{aligned}
         \dfrac{Pr(L=l+1|X=0)}{Pr(L=l|X=0)} = \dfrac{Pr(X=0|L=l+1)}{Pr(X=0|L=l)}
         = \dfrac{1 - Pr(X=1|L=l+1)}{1 - Pr(X=1|L=l)} \geq 1
    \end{aligned}
    \end{equation}
    \end{footnotesize}

    By combining Formula~\ref{eq:ap_p10},~\ref{eq:ap_p1}, and~\ref{eq:ap_p0}, the proposition is proved.
   
\end{proof}

\begin{figure}[t]
    \centering
    \includegraphics[width=3.1in]{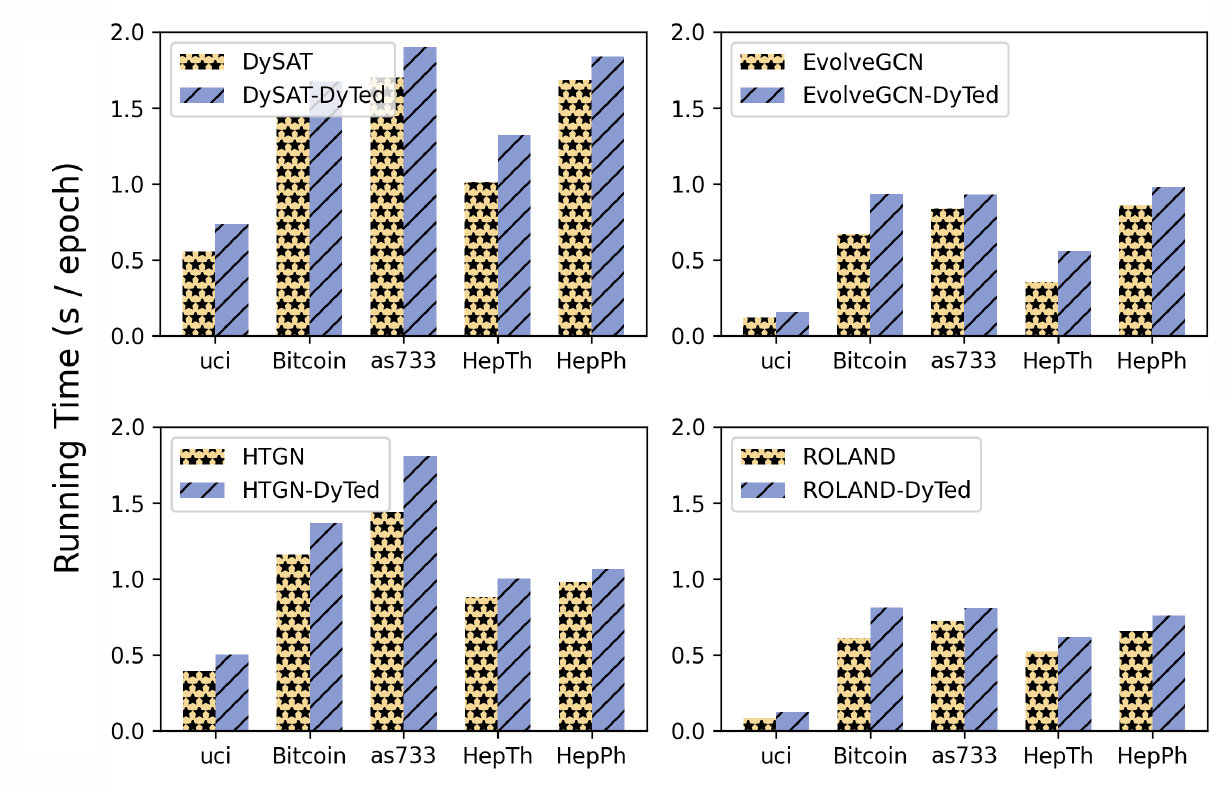}
    \caption{
        Running time of baselines and our framework. 
    }
    \label{fig:running}
\end{figure}

\begin{table}
\centering
\caption{Node classification for Pooling Representation}
\resizebox{3.4in}{!}{
    \begin{threeparttable}
    \begin{tabular}{lcccc}
    
    \toprule
    \multirow{2}[2]{*}{\textbf{Model}} & \multicolumn{2}{c}{\textbf{Assets}} & \multicolumn{2}{c}{\textbf{Financing Risk}}\\
                                         \cmidrule(r){2-3}                            \cmidrule(r){4-5}
                                       & \textbf{micro-F1} & \textbf{macro-F1}      & \textbf{micro-F1} & \textbf{macro-F1}\\
    \midrule
    \textbf{LSTMGCN}                   & 24.06 $\pm$ 0.46 & 16.37 $\pm$ 0.39 & 49.13 $\pm$ 0.32 & 29.56 $\pm$ 0.64\\
    \textbf{LSTMGCN-Pooling}\tnote{1}             & 23.07 $\pm$ 0.35 & 13.94 $\pm$ 0.13 & 47.19 $\pm$ 0.23 & 22.25 $\pm$ 0.13 \\
    \midrule
    \specialrule{0em}{0.5pt}{0.5pt}
    \midrule
    \textbf{DySAT}                      & 25.50 $\pm$ 0.58 & 13.06 $\pm$ 0.46 & 42.95 $\pm$ 2.46 & 25.03 $\pm$ 0.74 \\
    \textbf{DySAT-Pooling}                & 22.85 $\pm$ 0.66 & 12.76 $\pm$ 0.27 & 34.51 $\pm$ 1.22 & 26.13 $\pm$ 0.92\\
    \midrule
    \specialrule{0em}{0.5pt}{0.5pt}
    \midrule
    \textbf{EvolveGCN}                 & 24.39 $\pm$ 1.02 & 11.62 $\pm$ 0.17 & 43.76 $\pm$ 1.00 & 25.76 $\pm$ 0.40 \\
    \textbf{EvolveGCN-Pooling}           & 26.02 $\pm$ 2.23 & 8.76 $\pm$ 0.20 & 34.21 $\pm$ 0.15 & 16.99 $\pm$ 0.52 \\
    \midrule
    \specialrule{0em}{0.5pt}{0.5pt}
    \midrule
    \textbf{HTGN}                      & 26.70 $\pm$ 0.39 & 11.25 $\pm$ 0.71 & 50.13 $\pm$ 0.19 & 22.26 $\pm$ 0.06 \\
    \textbf{HTGN-Pooling}                 & 25.45 $\pm$ 0.67 & 8.66 $\pm$ 0.31 & 50.15 $\pm$ 0.22 & 21.36 $\pm$ 0.79\\
    \midrule
    \specialrule{0em}{0.5pt}{0.5pt}
    \midrule
    \textbf{ROLAND}                    & 28.08 $\pm$ 0.43 &  8.77 $\pm$ 0.10 & 49.34 $\pm$ 0.42 & 22.02 $\pm$ 0.12 \\
    \textbf{ROLAND-Pooling}              & 27.98 $\pm$ 0.13 & 8.71 $\pm$ 0.36 & 47.41 $\pm$ 0.93 & 21.31 $\pm$ 0.66 \\
    \bottomrule
    \end{tabular}
    \begin{tablenotes}
      \item[1]  The pooling of the representations over all snapshots of the baselines.
    \end{tablenotes}
    \end{threeparttable}
}

\label{tab:appendix_ncls_avg}%
\end{table}%

\begin{table*}[t]
\centering
\caption{Link prediction}
\resizebox{\textwidth}{!}{
    \begin{threeparttable}
    \begin{tabular}{lcccccccccc}
    
    \toprule
    \multirow{2}[2]{*}{\textbf{Model}} & \multicolumn{2}{c}{\textbf{Uci}} & \multicolumn{2}{c}{\textbf{Bitcoin}}        & \multicolumn{2}{c}{\textbf{as733}}    & \multicolumn{2}{c}{\textbf{HepTh}}   & \multicolumn{2}{c}{\textbf{HepPh}} \\
                                         \cmidrule(r){2-3}                            \cmidrule(r){4-5}                         \cmidrule(r){6-7}                        \cmidrule(r){8-9}                               \cmidrule(r){10-11}
                                       & \textbf{AUC} & \textbf{AP}      & \textbf{AUC} & \textbf{AP}   & \textbf{AUC} & \textbf{AP}  & \textbf{AUC} & \textbf{AP}         & \textbf{AUC} & \textbf{AP}\\
    \midrule
    \textbf{LSTMGCN-Time-invariant}\tnote{1}                   & 71.73 $\pm$ 0.68 & 69.05 $\pm$ 1.5 & 64.75 $\pm$ 0.48 & 64.71 $\pm$ 0.4 & 73.16 $\pm$ 0.13 & 72.56 $\pm$ 0.28 & 75.09 $\pm$ 0.71 & 74.89 $\pm$ 0.73 & 78.99 $\pm$ 0.72 & 78.74 $\pm$ 0.83\\
    \textbf{LSTMGCN-Time-varying}             & 70.08 $\pm$ 0.42 & 70.09 $\pm$ 1.33 & 75.1 $\pm$ 0.62 & 74.94 $\pm$ 0.61 & 74.53 $\pm$ 1.15 & 73.94 $\pm$ 1.27 & 81.02 $\pm$ 0.16 & 81.15 $\pm$ 0.14 & 78.91 $\pm$ 0.36 & 79.78 $\pm$ 0.39 \\
    \midrule
    \specialrule{0em}{0.5pt}{0.5pt}
    \midrule
    \textbf{DySAT-Time-invariant}                     & 81.84 $\pm$ 0.44 & 81.56 $\pm$ 0.39 & 74.48 $\pm$ 1.47 & 74.28 $\pm$ 1.46 & 77.48 $\pm$ 1.32 & 77.61 $\pm$ 1.31 & 71.43 $\pm$ 0.89 & 72.04 $\pm$ 0.87 & 61.65 $\pm$ 0.68 & 61.8 $\pm$ 0.72\\
    \textbf{DySAT-Time-varying}               & 80.16 $\pm$ 0.85 & 79.2 $\pm$ 0.65 & 76.27 $\pm$ 0.85 & 76.62 $\pm$ 0.81 & 72.77 $\pm$ 0.13 & 73.34 $\pm$ 0.13 & 73.31 $\pm$ 1.19 & 71.26 $\pm$ 1.23 & 77.4 $\pm$ 1.64 & 77.36 $\pm$ 1.66 \\
    \midrule
    \specialrule{0em}{0.5pt}{0.5pt}
    \midrule
    \textbf{EvolveGCN-Time-invariant}                 & 83.98 $\pm$ 0.87 & 89.11 $\pm$ 0.82 & 81.91 $\pm$ 0.46 & 82.02 $\pm$ 0.44 & 75.52 $\pm$ 0.35 & 73.5 $\pm$ 2.31 & 72.93 $\pm$ 1.91 & 73.42 $\pm$ 1.93 & 69.37 $\pm$ 1.76 & 69.46 $\pm$ 1.72\\
    \textbf{EvolveGCN-Time-varying}                  & 56.66 $\pm$ 0.47 & 57.54 $\pm$ 0.81 & 69.64 $\pm$ 1.53 & 69.73 $\pm$ 1.57 & 70.51 $\pm$ 1.82 & 70.93 $\pm$ 1.85 & 79.88 $\pm$ 1.45 & 79.59 $\pm$ 1.43 & 67.8 $\pm$ 2.04 & 68.01 $\pm$ 2.08\\
    \midrule
    \specialrule{0em}{0.5pt}{0.5pt}
    \midrule
    \textbf{HTGN-Time-invariant}                      & 91.57 $\pm$ 0.12 & 90.17 $\pm$ 0.38 & 84.33 $\pm$ 0.29 & 83.92 $\pm$ 0.2 & 76.51 $\pm$ 0.14 & 76.32 $\pm$ 0.29 & 72.59 $\pm$ 0.93 & 72.86 $\pm$ 0.97 & 72.65 $\pm$ 0.67 & 72.48 $\pm$ 0.62\\
    \textbf{HTGN-Time-varying}                & 80.73 $\pm$ 0.32 & 79.67 $\pm$ 0.36 & 87.01 $\pm$ 0.49 & 88.07 $\pm$ 0.48 & 72.14 $\pm$ 0.27 & 72.29 $\pm$ 0.39 & 75.56 $\pm$ 0.35 & 75.82 $\pm$ 0.35 & 70.6 $\pm$ 0.86 & 70.65 $\pm$ 0.86 \\
    \midrule
    \specialrule{0em}{0.5pt}{0.5pt}
    \midrule
    \textbf{ROLAND-Time-invariant}                     & 69.7 $\pm$ 1.12 & 67.37 $\pm$ 1.21 & 73.08 $\pm$ 2.01 & 73.06 $\pm$ 2.21 & 73.96 $\pm$ 0.26 & 74.25 $\pm$ 0.12 & 79.47 $\pm$ 0.21 & 79.53 $\pm$ 0.12 & 79.39 $\pm$ 0.37 & 78.78 $\pm$ 0.31 \\
    \textbf{ROLAND-Time-varying}              & 87.01 $\pm$ 0.93 & 87.21 $\pm$ 0.63 & 88.61 $\pm$ 0.05 & 88.62 $\pm$ 0.12 & 74.63 $\pm$ 0.35 & 74.11 $\pm$ 0.13 & 80.11 $\pm$ 0.6 & 79.67 $\pm$ 0.39 & 80.47 $\pm$ 0.51 & 79.97 $\pm$ 0.29 \\
    \bottomrule
   
    \end{tabular}
    \begin{tablenotes}
      \item[1] baseline-Time-invariant and baseline-Time-Varying are under our framework DyTed.
    \end{tablenotes}
    \end{threeparttable}
}

\label{tab:appendix_link}%
\end{table*}%

\begin{figure*}
    \centering
    \includegraphics[width=6.8in]{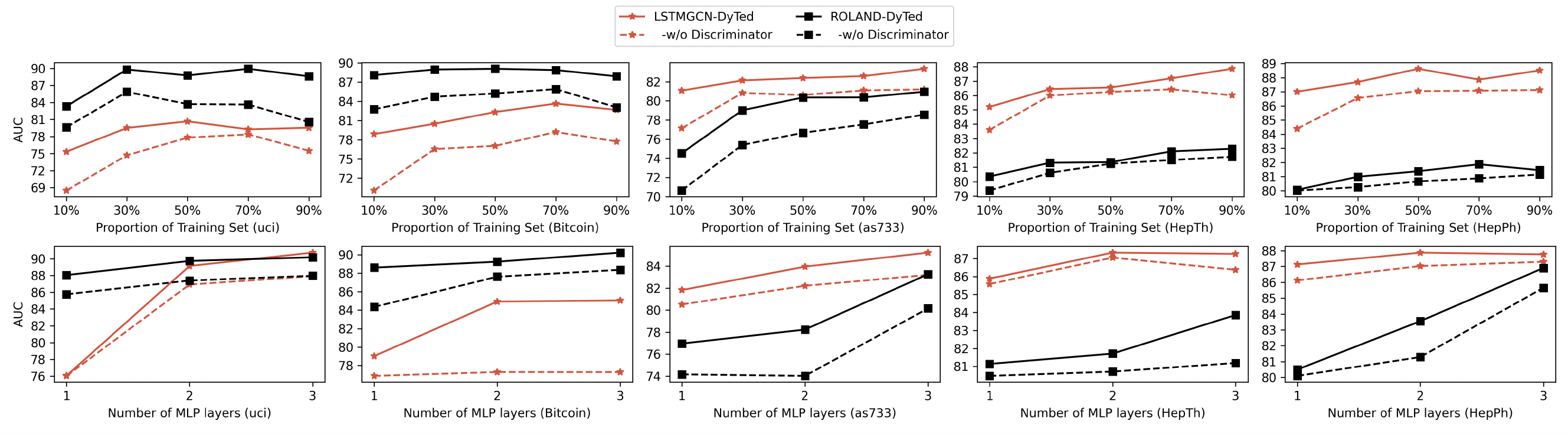}
    \caption{
        The benefits of disentanglement in terms of training resources for LSTMGCN and ROLAND.
    }
    \label{fig:appresources2}
\end{figure*}

\begin{figure*}
    \centering
    \includegraphics[width=6.8in]{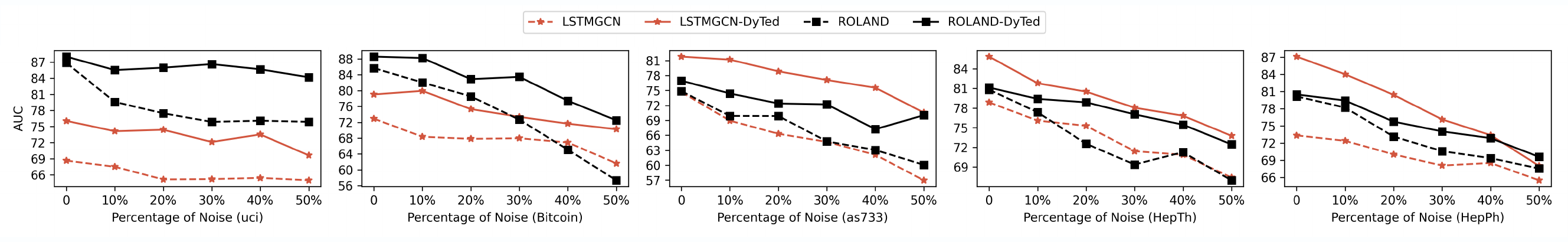}
    \caption{
        Performance against different noise rates for LSTMGCN and ROLAND.
    }
    \label{fig:approbustness2}
\end{figure*}

\begin{figure}
    \centering
    \includegraphics[width=3.4in]{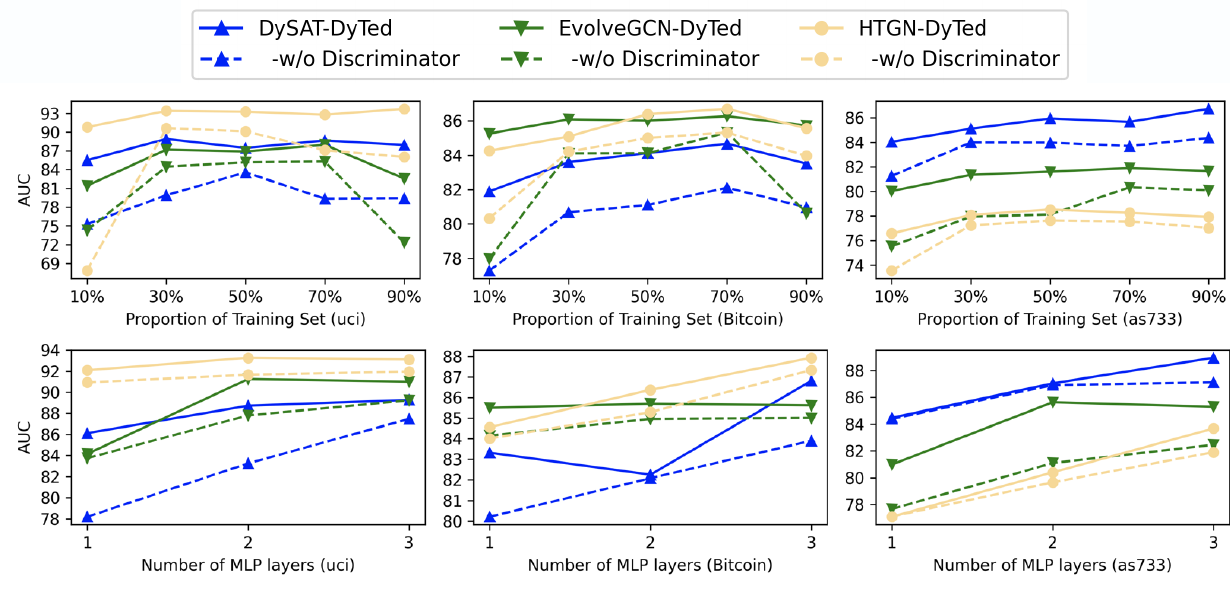}
    \caption{
        The benefits of disentanglement in terms of training resources.
    }
    \label{fig:appresources}
\end{figure}

\begin{figure}
    \centering
    \includegraphics[width=3.4in]{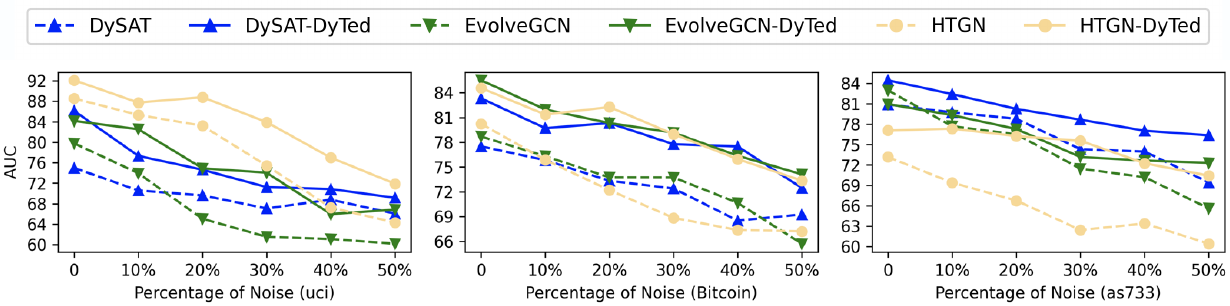}
    \caption{
        Performance against different noise rates.
    }
    \label{fig:approbustness}
\end{figure}

\subsection{Running Time Comparison}
\label{appendix:runtime}
Figure~\ref{fig:running} demonstrates the running time between baselines and our framework DyTed, which is comparable in order of magnitude.

\subsection{Performance of Pooling Representation}
\label{appendix:poolrep}
Table~\ref{tab:appendix_ncls_avg} demonstrates the performance of pooling representations among all snapshots for node classification, showing a worse $F1$ than the representation of the last snapshot. (Due to the space limitation, we only show the results on two labels.)

\subsection{Supplement}
\label{appendix:supp}
We have included Table~\ref{tab:appendix_link} to supplement Table~\ref{tab:l-prediction} in Section~\ref{sec:performance}, Figure~\ref{fig:appresources2}, Figure~\ref{fig:approbustness2}, Figure~\ref{fig:appresources}, and Figure~\ref{fig:approbustness} to supplement Figure~\ref{fig:trainig_res} and Figure~\ref{fig:robustness} in Section~\ref{sec:benefit}. These additional tables support the same conclusions.

\end{document}